\documentclass[conference]{IEEEtran}

\usepackage[utf8]{inputenc}
\usepackage[T1]{fontenc}

\usepackage{booktabs}
\usepackage{microtype}
\usepackage[caption=false,font=footnotesize]{subfig}
\usepackage[hyphens]{url}
\usepackage{cite}

\usepackage{makecell}

\let\labelindent\relax
\usepackage[inline]{enumitem}

\usepackage{tikz}

\usepackage[colorlinks,allcolors=black,plainpages=false,pdfusetitle,pdfauthor={Victor Le Pochat, Tom Van Goethem, Samaneh Tajalizadehkhoob, Maciej Korczyński, Wouter Joosen}]{hyperref}

\graphicspath{{plots/}{./}}

\newcommand{\etal}{et\,\,al.}
\newcommand{\listname}{\textsc{Tranco}}

\begin{document}
	
\title{\listname: A Research-Oriented Top Sites Ranking\\Hardened Against Manipulation}

\author{\IEEEauthorblockN{Victor Le Pochat\IEEEauthorrefmark{1}, Tom Van Goethem\IEEEauthorrefmark{1}, Samaneh Tajalizadehkhoob\IEEEauthorrefmark{2}, Maciej Korczy\'{n}ski\IEEEauthorrefmark{3}, Wouter Joosen\IEEEauthorrefmark{1}\vspace{0.5em}}
	\IEEEauthorblockA{
		\begin{tabular}{ccc}
			\begin{tabular}{@{}c@{}}
				\IEEEauthorrefmark{1}imec-DistriNet, KU Leuven\\\{firstname.lastname\}@cs.kuleuven.be
			\end{tabular} & \begin{tabular}{@{}cc@{}}
				\IEEEauthorrefmark{2}Delft University of Technology\\S.T.Tajalizadehkhoob@tudelft.nl
			\end{tabular}& \begin{tabular}{@{}c@{}}
				\IEEEauthorrefmark{3}Grenoble Alps University\\maciej.korczynski@univ-grenoble-alpes.fr
			\end{tabular}
		\end{tabular}
	}
}

\IEEEoverridecommandlockouts
\makeatletter\def\@IEEEpubidpullup{6.5\baselineskip}\makeatother
\IEEEpubid{\parbox{\columnwidth}{
    Network and Distributed Systems Security (NDSS) Symposium 2019\\
    24-27 February 2019, San Diego, CA, USA\\
    ISBN 1-891562-55-X\\
    https://dx.doi.org/10.14722/ndss.2019.23386\\
    www.ndss-symposium.org
}
\hspace{\columnsep}\makebox[\columnwidth]{}}

\maketitle

\begin{abstract}
In order to evaluate the prevalence of security and privacy practices on a representative sample of the Web, researchers rely on website popularity rankings such as the Alexa list. While the validity and representativeness of these rankings are rarely questioned, our findings show the contrary: we show for four main rankings how their inherent properties (similarity, stability, representativeness, responsiveness and benignness) affect their composition and therefore potentially skew the conclusions made in studies. Moreover, we find that it is trivial for an adversary to manipulate the composition of these lists. We are the first to empirically validate that the ranks of domains in each of the lists are easily altered, in the case of Alexa through as little as a single HTTP request. This allows adversaries to manipulate rankings on a large scale and insert malicious domains into whitelists or bend the outcome of research studies to their will.
To overcome the limitations of such rankings, we propose improvements to reduce the fluctuations in list composition and guarantee better defenses against manipulation. To allow the research community to work with reliable and reproducible rankings, we provide \listname, an improved ranking that we offer through an online service available at \url{https://tranco-list.eu}.
\end{abstract}

\section{Introduction}
Researchers and security analysts frequently study a selection of popular sites, such as for measuring the prevalence of security issues or as an evaluation set of available and often used domain names, as these are purported to reflect real-world usage. The most well known and widely used list in research studies is that of Alexa, with researchers' reliance on this commercial list being accentuated by their concern when it was momentarily taken offline in November 2016~\cite{Twitter1}. However, several companies provide alternative rankings based on Internet usage data collected through various channels~\cite{Napoli2014}: a panel of users whose visits are logged, tracking code placed on websites and traffic captured by intermediaries such as ISPs. 

We found that 133 top-tier studies over the past four years based their experiments and conclusions on the data from these rankings. Their validity and by extension that of the research that relies on them, should however be questioned: the methods behind the rankings are not fully disclosed, and commercial interests may prevail in their composition. Moreover, the providers only have access to a limited userbase that may be skewed towards e.g.\ certain user groups or geographic regions. Even though most providers declare that the data is processed to remove such statistical biases, the lack of exact details makes it impossible for researchers to assess the potential impact of these lists on their results and conclusions.

In this paper, we show that the four main popularity rankings (Alexa, Cisco Umbrella, Majestic and Quantcast) exhibit significant problems for usage in research. The rankings hardly agree on the popularity of any domain, and the Umbrella and especially the Alexa lists see a significant turnover even on consecutive days; for Alexa, this is the result of an unannounced and previously unknown change in averaging approach. All lists include non-representative and even malicious sites, which is especially dangerous considering the widespread use of these rankings as whitelists. Overall, these flaws can cause the choice for a particular ranking to severely skew measurements of vulnerabilities or secure practices.

Moreover, we are the first to empirically prove that pitfalls in these rankings leave them vulnerable to one of our newly introduced manipulation techniques. These techniques have a surprisingly low cost, starting from a single HTTP request for Alexa, and can therefore be used to affect the rank of thousands of domains at once on a substantial level: we estimate that the top 10\,000 can easily be reached. The incentives of adversaries to alter the composition of these lists, both for single domains due to the practice of whitelisting popular domains, and on a larger scale to influence research and its impact outside academia, make this manipulation particularly valuable.

Finally, there is still a need for researchers to study popular domains, so they would therefore benefit from a list that avoids biases in its inherent properties and is more resilient to manipulation, and that is easily retrieved for future reference. To this extent, we propose improvements to current rankings in terms of stability over time, representativeness and hardening against manipulation. We create \listname, a new ranking that is made available and archived through an accompanying online service at \url{https://tranco-list.eu}, in order to enhance the reproducibility of studies that rely on them. The community can therefore continue to study the security of popular domains while ensuring valid and verifiable research.

\medskip \noindent In summary, we make the following contributions:
\begin{itemize}
	\item We describe how the main rankings can negatively affect security research, e.g.\ half of the Alexa list changes every day and the Umbrella list only has 49\% real sites, as well as security implementations, e.g.\ the Majestic list contains 2\,162 malicious domains despite being used as a whitelist.
	\item We classify how 133 recent security studies rely on these rankings, in particular Alexa, and show how adversaries could exploit the rankings to bias these studies.
	\item We show that for each list there exists at least one technique to manipulate it on a large scale, as e.g.\ only one HTTP request suffices to enter the widely used Alexa top million. We empirically validate that reaching a rank as good as 28\,798 is easily achieved.
	\item Motivated by the discovered limitations of the widely-used lists, we propose \listname, an alternative list that is more appropriate for research, as it varies only by 0.6\% daily and requires at least the quadrupled manipulation effort to achieve the same rank as in existing lists.
\end{itemize}

\section{Methodology of top websites rankings}\label{sec:lists}

Multiple commercial providers publish rankings of popular domains that they compose using a variety of methods. For Alexa, Cisco Umbrella, Majestic and Quantcast, the four lists that are available for free in an easily parsed format and that are regularly updated, we discuss what is known on how they obtain their data, what metric they use to rank domains and which potential biases or shortcomings are present. We base our discussion mainly on the documentation available from these providers; many components of their rankings are proprietary and could therefore not be included.

We do not consider any lists that require payment, such as SimilarWeb\footnote{\url{https://www.similarweb.com/top-websites}}, as their cost (especially for longitudinal studies) and potential usage restrictions make them less likely to be used in a research context. We also disregard lists that would require scraping, such as Netcraft\footnote{https://toolbar.netcraft.com/stats/topsites}, as these do not carry the same consent of their provider implied by making the list available in a machine-readable format. Finally, Statvoo's list\footnote{\url{https://statvoo.com/dl/top-1million-sites.csv.zip}} seemingly meets our criteria. However, we found it to be a copy of Alexa's list of November 23, 2016, having never been updated since; we therefore do not consider it in our analysis.

\subsection{Alexa} Alexa, a subsidiary of Amazon, publishes a daily updated list\footnote{\url{https://s3.amazonaws.com/alexa-static/top-1m.csv.zip}} consisting of one million websites since December 2008~\cite{Alexa5}. Usually only pay-level domains\footnote{A \emph{pay-level domain} (PLD) refers to a domain name that a consumer or business can directly register, and consists of a subdomain of a \emph{public suffix} or \emph{effective top-level domain} (e.g.\ \texttt{.com} but also \texttt{.co.uk}).} are ranked, except for subdomains of certain sites that provide `personal home pages or blogs'~\cite{Alexa1} (e.g.\ \url{tmall.com}, \url{wordpress.com}). In November 2016, Alexa briefly took down the free CSV file with the list~\cite{Twitter1}. The file has since been available again~\cite{Twitter2} and is still updated daily; however, it is no longer linked to from Alexa's main website, instead referring users to the paid `Alexa Top Sites' service on Amazon Web Services~\cite{Alexa6}.

The ranks calculated by Alexa are based on traffic data from a ``global data panel'', with domains being ranked on a proprietary measure of unique visitors and page views, where one visitor can have at most one page view count towards the page views of a URL~\cite{Alexa2}. Alexa states that it applies ``data normalization'' to account for biases in their user panel~\cite{Alexa1}. 

The panel is claimed to consist of millions of users, who have installed one of ``many different'' browser extensions that include Alexa's measurement code~\cite{Alexa8}. However, through a crawl of all available extensions for Google Chrome and Firefox, we found only Alexa's own extension (``Alexa Traffic Rank'') to report traffic data. Moreover, this extension is only available for the desktop version of these two browsers. Chrome's extension is reported to have around 570\,000 users~\cite{alexaextension}; no user statistics are known for Firefox, but extrapolation based on browser usage suggests at most one million users for two extensions, far less than Alexa's claim. 

In addition, sites can install an `Alexa Certify' tracking script that collects traffic data for all visitors; the rank can then be based on these actual traffic counts instead of on estimates from the extension~\cite{Alexa1}. This service is estimated to be used by 1.06\% of the top one million and 4\% of the top 10\,000~\cite{BuildWithCertify}. 

The rank shown in a domain's profile on Alexa's website is based on data over three months, while in 2016 they stated that the downloadable list was based on data over one month~\cite{Alexa3a}. This statement was removed after the brief takedown of this list~\cite{Alexa3b}, but the same period was seemingly retained. However, as we derive in Section~\ref{sub:stability}, since January 30, 2018 the list is based on data for one day; this was confirmed to us by Alexa but was otherwise unannounced.

Alexa's data collection method leads to a focus on sites that are visited in the top-level browsing context of a web browser (i.e.\ HTTP traffic). They also indicate that ranks worse than 100\,000 are not statistically meaningful, and that for these sites small changes in measured traffic may cause large rank changes~\cite{Alexa1}, negatively affecting the stability of the list. 

\subsection{Cisco Umbrella} Cisco Umbrella publishes a daily updated list\footnote{\url{https://s3-us-west-1.amazonaws.com/umbrella-static/top-1m.csv.zip}} consisting of one million entries since December 2016~\cite{Umbrella2}. Any domain name may be included, with it being ranked on the aggregated traffic counts of itself and all its subdomains.

The ranks calculated by Cisco Umbrella are based on DNS traffic to its two DNS resolvers (marketed as OpenDNS), claimed to amount to over 100 billion daily requests from 65 million users~\cite{Umbrella2}. Domains are ranked on the number of unique IPs issuing DNS queries for them~\cite{Umbrella2}. Not all traffic is said to be used: instead the DNS data is sampled and `data normalization methodologies' are applied to reduce biases~\cite{Umbrella1}, taking the distribution of client IPs into account~\cite{Umbrella3}. Umbrella's data collection method means that non-browser-based traffic is also accounted for. A side-effect is that invalid domains are also included (e.g.\ internal domains such as \url{*.ec2.internal} for Amazon EC2 instances, or typos such as \url{google.conm}).

\subsection{Majestic} Majestic publishes the daily updated `Majestic Million' list consisting of one million websites\footnote{\url{http://downloads.majestic.com/majestic_million.csv}} since October 2012~\cite{Majestic2}. The list comprises mostly pay-level domains, but includes subdomains for certain very popular sites (e.g.\ \url{plus.google.com}, \url{en.wikipedia.org}).

The ranks calculated by Majestic are based on backlinks to websites, obtained by a crawl of around 450~billion URLs over 120~days, changed from 90 days on April 12, 2018~\cite{Majestic1, Majestic5}. Sites are ranked on the number of class C (IPv4 /24) subnets that refer to the site at least once~\cite{Majestic2}. Majestic's data collection method means only domains linked to from other websites are considered, implying a bias towards browser-based traffic, however without counting actual page visits. Similarly to search engines, the completeness of their data is affected by how their crawler discovers websites.

\subsection{Quantcast} Quantcast publishes a list\footnote{\url{https://ak.quantcast.com/quantcast-top-sites.zip}} of the websites visited the most in the United States since mid 2007~\cite{Quantcast3}. The size of the list varies daily, but usually was around 520,000 mostly pay-level domains; subdomains reflect sites that publish user content (e.g.\ \url{blogspot.com}, \url{github.io}). The list also includes `hidden profiles', where sites are ranked but the domain is hidden.

The ranks calculated by Quantcast are based on the number of people visiting a site within the previous month, and comprises `quantified' sites where Quantcast directly measures traffic through a tracking script as well as sites where Quantcast estimates traffic based on data from `ISPs and toolbar providers'~\cite{Quantcast1}. These estimates are only calculated for traffic in the United States, with only quantified sites being ranked in other countries; the list of top sites also only considers US traffic. Moreover, while quantified sites see their visit count updated daily, estimated counts are only updated monthly~\cite{Quantcast2}, which may inflate the stability of the list. Before November 14, 2018, quantified sites made up around 10\% of the full (US) list. However, since then Quantcast seems to have stopped ranking almost any estimated domains, therefore reducing the list size to around 40\,000.\section{Quantitative comparison}\label{sec:comparison}
Ideally, the domain rankings would perfectly reflect the popularity of websites, free from any biases. However, the providers of domain rankings do not have access to complete Internet usage data and use a variety of largely undisclosed data collection and processing methods to determine the metric on which they rank websites. This may lead to differences between the lists and potential `hidden' factors influencing the rankings: the choice of list can then critically affect e.g.\ studies that measure the prevalence of security practices or vulnerabilities. We compare the four main lists over time in order to assess the breadth and impact of these differences.

Certain properties may reflect how accurately Internet usage is measured and may be (more or less) desired when using the lists for security research. We consider five properties in our comparison: \begin{enumerate*}
	\item \emph{similarity} or the agreement on the set of popular domains,
	\item \emph{stability} or the rank changes over time,
	\item \emph{representativeness} or the reflection of popularity across the web,
	\item \emph{responsiveness} or the availability of the listed websites, and
	\item \emph{benignness} or the lack of malicious domains
\end{enumerate*}.

To quantitatively assess these properties, we use the lists obtained between January~1 and November~30, 2018, referring to the date when the list would be downloaded; the data used by the provider to compile the list may be older. In addition, we crawled the sites on the four lists as downloaded on May~11, 2018 at 13:00 UTC from a distributed crawler setup of 10 machines with 4 CPU cores and 8~GB RAM in our European university network, using Ubuntu 16.04 with Chromium version 66.0.3359.181 in headless mode.

\subsection{Similarity}
\begin{figure}
	\centering
	\includegraphics[width=0.95\columnwidth]{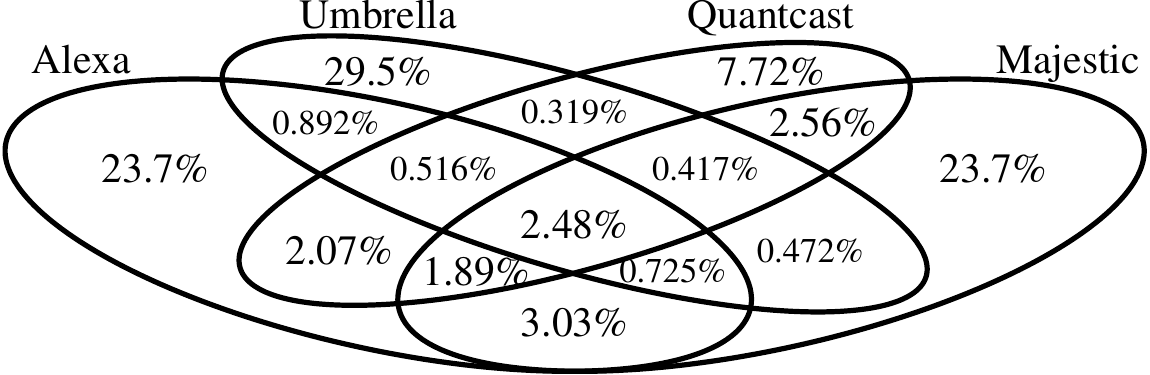}
	\caption{The average daily intersections between the lists of the four providers from January 30, 2018 to November 13, 2018.}
	\label{fig:intersections}
\end{figure}

Figure~\ref{fig:intersections} shows the average number of sites that the rankings agree upon per day; there is little variance over time. The four lists combined contain around 2.82~million sites, but agree only on around 70\,000 sites. Using the rank-biased overlap (RBO)~\cite{Webber2010a}, a similarity measure that can be parameterized to give a higher weight to better ranks, we see that the lists of Alexa, Majestic and Quantcast are the most similar to each other. However, even when heavily weighting the top 100, the RBO remains low between 24\% and 33\%. Umbrella's full list is most dissimilar to the others, with an RBO of between 4.5\% and 15.5\%. However, this is to be expected as Umbrella includes subdomains: when ranking only pay-level domains, the RBO with the other lists reaches around 30\% as well. Finally, Quantcast's removal of non-quantified sites after November 14, 2018 causes a significant drop in RBO to less than 5.5\%, with no overlap of the top~10: many very popular domains are not quantified and are therefore now missing from Quantcast's list. 

The small overlaps signify that there is no agreement on which sites are the most popular. This means that switching lists yields a significantly different set of domains that can e.g.\ change how prevalent certain web trackers seem to be~\cite{Englehardt2016}.

\subsection{Stability}\label{sub:stability}

\begin{figure}
	\centering
	\includegraphics[width=1\linewidth]{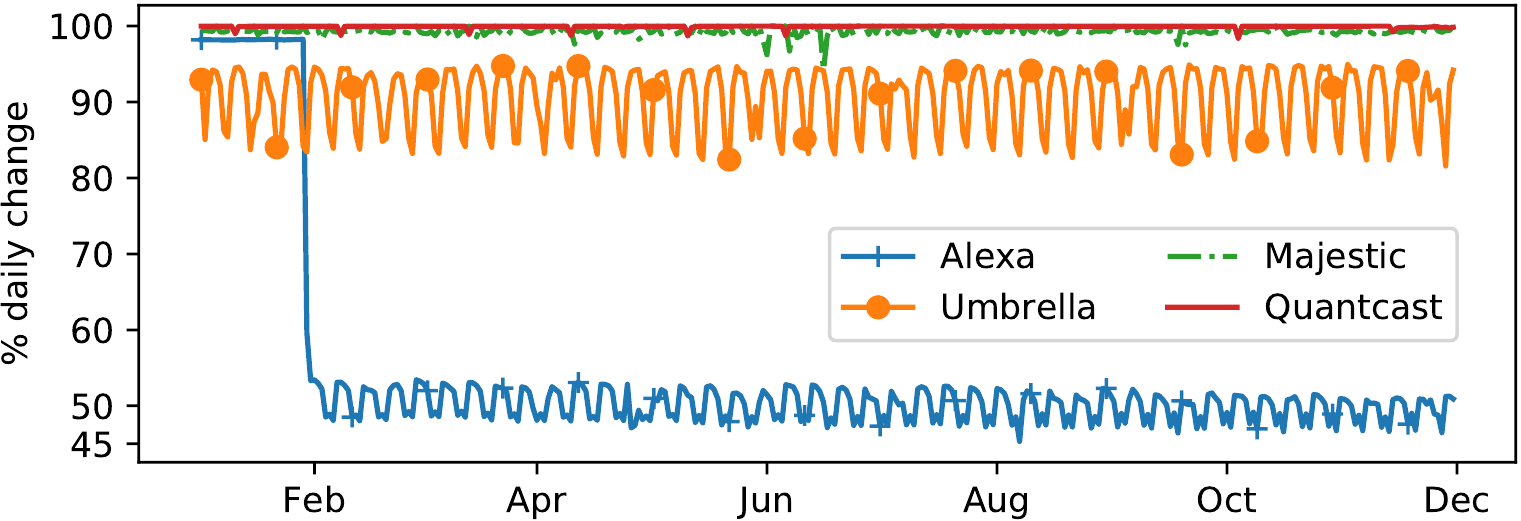}
	\caption{The intersection percentage between each provider's lists for two consecutive days.}
	\label{fig:stability-intersections}
\end{figure}

From the intersections between each provider's lists for two consecutive days, shown in Figure~\ref{fig:stability-intersections}, we see that Majestic's and Quantcast's lists are the most stable, usually changing at most 1\% per day, while for Umbrella's list this climbs to on average 10\%. Until January 30, 2018, Alexa's list was almost as stable as Majestic's or Quantcast's. However, since then stability has dropped sharply, with around half of the top million changing every day, due to Alexa's change to a one day average. There exists a trade-off in the desired level of stability: a very stable list provides a reusable set of domains, but may therefore incorrectly represent sites that suddenly gain or lose popularity. A volatile list however may introduce large variations in the results of longitudinal studies.

\subsection{Representativeness}\label{sub:representativeness}
\begin{figure}
	\centering
	\includegraphics[width=1\linewidth]{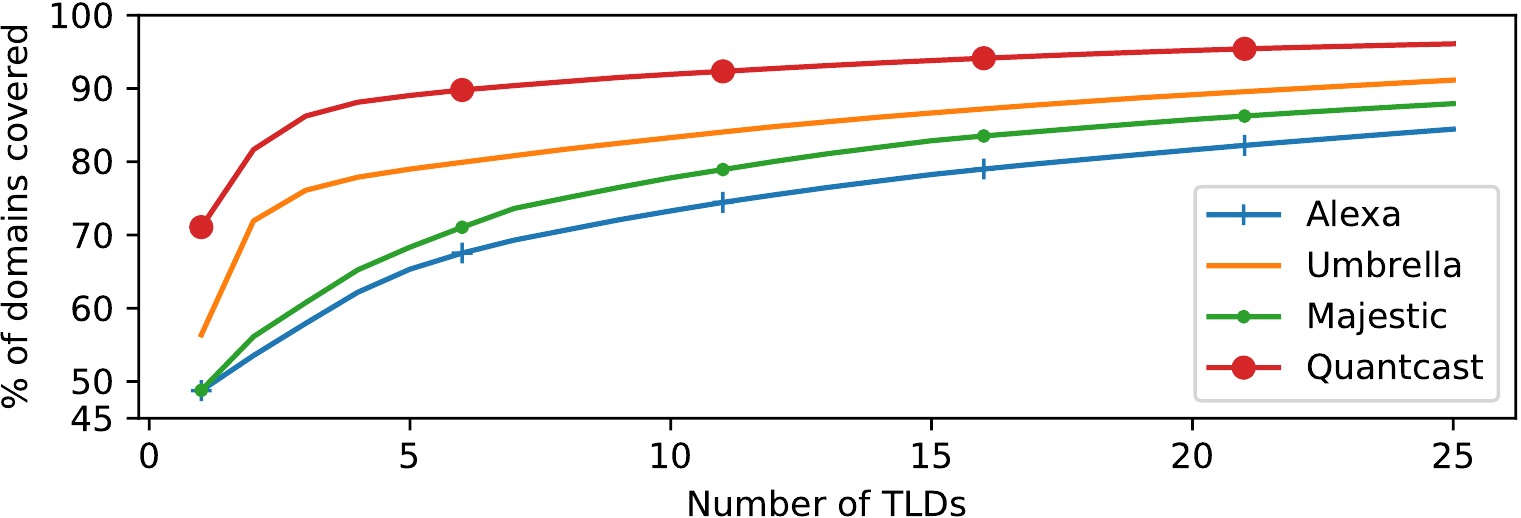}
	\caption{The cumulative distribution function of TLD usage across the lists.}
	\label{fig:tlds-cdf}
\end{figure}

Sites are mainly distributed over a few top-level domains, with Figure~\ref{fig:tlds-cdf} showing that 10 TLDs capture more than 73\% of every list. The .com TLD is by far the most popular, at almost half of Alexa's and Majestic's list and 71\% of Quantcast's list; .net, .org and .ru are used most often by other sites. One notable outlier is the .jobs TLD: while for the other lists it does not figure in the top~10 TLDs, it is the fourth most popular TLD for Quantcast. Most of these sites can be traced to DirectEmployers, with thousands of lowly ranked domains. This serves as an example of one entity controlling a large part of a ranking, potentially giving them a large influence in research results.

We use the autonomous system to determine the entities that host the ranked domains. Google hosts the most websites within the top 10 and 100 sites, at between 15\% and 40\% except for Quantcast at 4\%: for Alexa these are the localized versions, for the other lists these are subdomains. For the full lists, large content delivery networks dominate, with Cloudflare being the top network hosting up to 10\% of sites across all lists.
This shows that one or a few entities may be predominantly represented in the set of domains used in a study and that therefore care should be taken when considering the wider implications of its results.

\subsection{Responsiveness}

\begin{figure}
	\centering
	\includegraphics[width=1\linewidth]{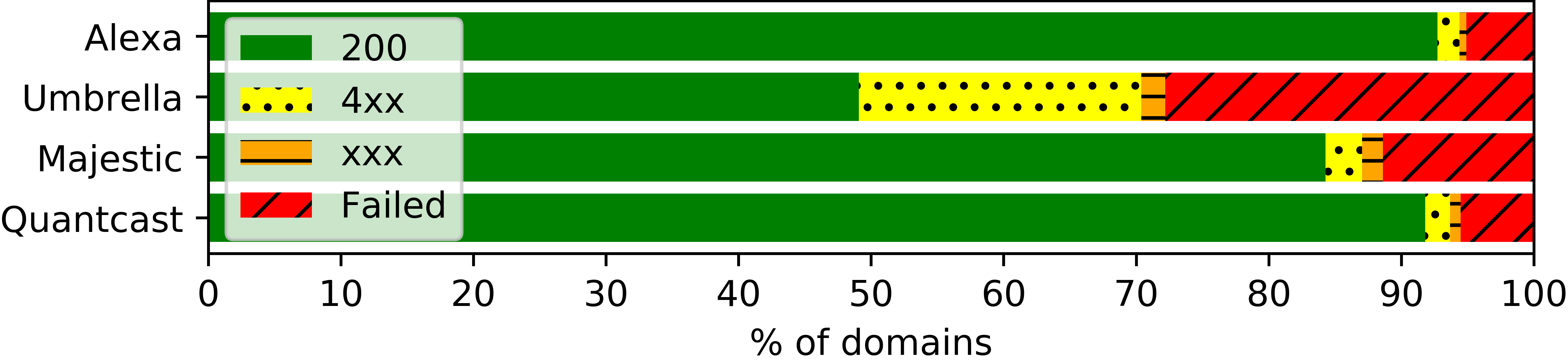}
	\caption{The responsiveness and reported HTTP status code across the lists.}
	\label{fig:responsiveness}
\end{figure}

Figure~\ref{fig:responsiveness} shows the HTTP status code reported for the root pages of the domains in the four lists. 5\% of Alexa's and Quantcast's list and 11\% of Majestic's list could not be reached. For Umbrella, this jumps to 28\%; moreover only 49\% responded with status code 200, and 30\% reported a server error. Most errors were due to name resolution failure, as invalid or unconfigured (sub)domains are not filtered out. 

Of the reachable sites, 3\% for Alexa and Quantcast, 8.7\% for Majestic and 26\% for Umbrella serve a page smaller than 512 bytes on their root page, based on its download size as reported by the browser instance. As such pages often appear empty to the user or only produce an error, this indicates that they may not contain any useful content, even though they are claimed to be regularly visited by real users. Unavailable sites and those without content do not represent real sites and may therefore skew e.g.\ averages of third-party script inclusion counts~\cite{Nikiforakis2012}, as these sites will be counted as having zero inclusions.

\subsection{Benignness}
Malicious campaigns may target popular domains to extend the reach of their attack, or use a common domain as a point of contact, leading to it being picked up as `popular'. While it is not the responsibility of ranking providers to remove malicious domains, popular sites are often assumed to be trustworthy, as evidenced by the practice of whitelisting them~\cite{Quad9Majestic} or, as we show in Section~\ref{sub:survey}, their usage in security research as the benign test set for classifiers.

Table~\ref{tbl:safebrowsing} lists the number of domains flagged on May 31, 2018 by Google Safe Browsing, used among others by Chrome and Firefox to automatically warn users when they visit dangerous sites~\cite{SafeBrowsing}. At 0.22\% of its list, Majestic has the most sites that are flagged as potentially harmful (in particular as malware sites), but all lists rank at least some malicious domains. In Alexa's top 10\,000, 4 sites are flagged as performing social engineering (e.g.\ phishing), while 1 site in Majestic's top 10\,000 serves unwanted software. The presence of these sites in Alexa's and Quantcast's list is particularly striking, as users would have to actively ignore the browser warning in order to trigger data reporting for Alexa's extension or the tracking scripts.

\begin{table}[t]
	\centering
	\caption{Presence of domains in the four rankings on Google's Safe Browsing list on May 31, 2018.}
	\label{tbl:safebrowsing}
	\resizebox{\linewidth}{!}{
		\begin{tabular}{lccccccccccc}
			\toprule
			&	\multicolumn{2}{c}{Malware} & \multicolumn{3}{c}{Social Engineering} & \multicolumn{3}{c}{Unwanted software} & \multicolumn{2}{c}{\makecell{Potentially\\ harmful\\ application}}& Total \\
			
			\cmidrule(r){2-3}\cmidrule(lr){4-6}\cmidrule(lr){7-9}\cmidrule(lr){10-11}
			&100K&Full&10K&100K&Full&10K&100K&Full&100K&Full&\\ \midrule
			Alexa&32 & 98 & 4 & 85 & 345 & 0 & 15 & 104 & 0 & 0 & 547\\
			Umbrella&11 & 326 & 0 & 3 & 393 & 0 & 23 & 232 & 4 & 60 & 1011\\
			Majestic&130 & 1676 & 0 & 23 & 359 & 1 & 9 & 79 & 9 & 48 & 2162\\
			Quantcast&3 & 76 & 0 & 4 & 105 & 0 & 4 & 41 & 0 & 2 & 224\\
			
			\bottomrule
			
		\end{tabular}
	}
\end{table}

Given the presence of malicious domains on these lists, the practice of whitelisting popular domains is particularly dangerous. Some security analysis tools whitelist sites on Alexa's list~\cite{NETRESEC, Mertens2017}. Moreover, Quad9's DNS-based blocking service whitelists all domains on Majestic's list~\cite{Quad9Majestic}, exposing its users to ranked malicious domains. As Quad9's users expect harmful domains to be blocked, they will be even more under the impression that the site is safe to browse; this makes the manipulation of the list very interesting to attackers.

\section{Usage in security research}
Whenever security issues are being investigated, researchers may want to evaluate their impact on real-world domains. For these purposes, security studies often use and reference the top sites rankings. The validity and representativeness of these rankings therefore directly affects their results, and any biases may prohibit correct conclusions being made. Moreover, if forged domains could be entered into these lists, an adversary can control research findings in order to advance their own goals and interests. 

\subsection{Survey and classification of list usage}\label{sub:survey}

To assess how security studies use these top sites rankings, we surveyed the papers from the main tracks of the four main academic security conferences (CCS, NDSS, S\&P, USENIX Security) from 2015 to 2018; we select these venues as they are considered top-tier and cover general security topics. We classify these papers according to four purposes for the lists: \emph{prevalence} if the rankings are used to declare the proportion of sites affected by an issue; \emph{evaluation} if a set of popular domains serves to test an attack or defense, e.g.\ for evaluating Tor fingerprinting~\cite{Rimmer2018}; \emph{whitelist} if the lists are seen as a source of benign websites, e.g.\ for use in a classifier~\cite{Vissers2015a}; \emph{ranking} if the exact ranks of sites are mentioned or used (e.g.\ to estimate website traffic~\cite{Englehardt2016}) or if sites are divided into bins according to their rank.

Alexa is by far the most popular list used in recent security studies, with 133 papers using the list for at least one purpose. Table~\ref{tbl:alexa-papers} shows the number of papers per category and per subset of the list that was used. The Alexa list is mostly used for measuring the prevalence of issues or as an evaluation set of popular domains. For the former purpose as well as for whitelisting and ranking or binning, the full list is usually used, while for evaluation sets, the subset size varies more widely. 
Three papers from these conferences also used another ranking, always in tandem with the Alexa list~\cite{Borgolte2015,Zimmeck2017,Zhang2018}.

\begin{table}
	\centering
	\caption{Categorization of recent security studies using the Alexa ranking. One study may appear in multiple categories.}
	\label{tbl:alexa-papers}
	\resizebox{\linewidth}{!}{	
		\begin{tabular}{lccccccccc}
			\toprule
			& \multicolumn{8}{c}{Subset studied} & \\
			\cmidrule{2-9}
			Purpose    & 10 & 100 & 500 & 1K & 10K & 100K & 1M & Other & Total \\
			\midrule
			Prevalence & 1  & 6   & 8   & 9  & 16  & 7    & 32 & 13    & 63 \\
			Evaluation & 7  & 16  & 14  & 10 & 9   & 3    & 14 & 28    & 71 \\
			Whitelist  & 0  & 2   & 1   & 4  & 3   & 2    & 11 & 6     & 19 \\
			Ranking    & 0  & 1   & 3   & 3  & 2   & 4    & 15 & 7     & 28 \\
			\midrule
			Total      & 8  & 20  & 18  & 18 & 23  & 9    & 45 & 36    & 133 \\
			\bottomrule
		\end{tabular}
	}
\end{table}

Most studies lack any comment on when the list was downloaded, when the websites on the lists were visited and what proportion was actually reachable. This hampers reproducibility of these studies, especially given the daily changes in list compositions and ranks.

Two papers commented on the methods of the rankings. Juba \etal~\cite{Juba2015} mention the rankings being ``representative of true traffic numbers in a coarse grained sense''. Felt \etal~\cite{Felt2017} mention the ``substantial churn'' of Alexa's list and the unavailability of sites, and express caution in characterizing all its sites as popular. However, in general the studies do not question the validity of the rankings, even though they have properties that can significantly affect their conclusions, and as we will show are vulnerable to manipulation. 

\subsection{Influence on security studies}\label{sub:influence}
\subsubsection{Incentives}
Given the increasing interest in cybersecurity within our society, the results of security research have an impact beyond academia. News outlets increasingly report on security vulnerabilities, often mentioning their prevalence or affected high-profile entities~\cite{LogjamWSJ, LogjamArs, STARTTLSArs, HTTPSArs }. Meanwhile, policy-makers and governments rely on these studies to evaluate secure practices and implement appropriate policies~\cite{Aydin2017, Doty2018}; e.g.\ Mozilla in part decided to delay distrusting Symantec certificates based on a measurement across Umbrella's list~\cite{MozillaSymantec}.

Malicious actors may therefore risk exposure to a wider audience, while their practices may trigger policy changes, yielding them an incentive to directly influence security studies. Invernizzi \etal~\cite{Invernizzi2016} discovered that blacklists sold on underground markets contain IP addresses of academic institutions as well as security companies and researchers, illustrating that adversaries already actively try to prevent detection by researchers. As we showed, security studies often rely on popularity rankings, so pitfalls in the methods of these rankings that expose them to targeted manipulation open up another opportunity for adversaries to affect security research. 
The way in which an adversary may want to influence rankings, and therefore the research dependent upon them, varies according to their incentives. They may want to promote domains into the lists, making them be perceived as benign and then execute malicious practices through them.  Alternatively, they can promote other domains to hide their own malicious domains from the lists. Finally, they can intelligently combine both techniques to alter comparisons of security properties for websites of different entities.

\subsubsection{Case study}
The issue of online tracking and fingerprinting has been studied on multiple occasions for Alexa's top one million~\cite{Nikiforakis2012, Libert2015, Englehardt2016, Kumar2017, Libert2018}. Users may want to avoid organizations that perform widespread or invasive tracking, and therefore have an interest in new tracking mechanisms and/or specific trackers being found or named by these studies, e.g.\ to include them in blocklists. The trackers therefore have an incentive to avoid detection by not figuring among the domains being studied, e.g.\ by pushing these out of the popularity ranking used to provide the set of investigated domains. 

We quantify the effort required to manipulate a ranking and therefore alter findings for the measurements of fingerprinting prevalence by Acar \etal~\cite{Acar2014} and Englehardt and Narayanan~\cite{Englehardt2016} on Alexa's top 100\,000 and top one million respectively. These studies published data on which domains included which scripts, including the Alexa rank. We calculate how many domains minimally need to be moved up in order to push out the websites using a particular tracking provider.

\begin{figure}[t]
	\centering
	\includegraphics[width=\linewidth]{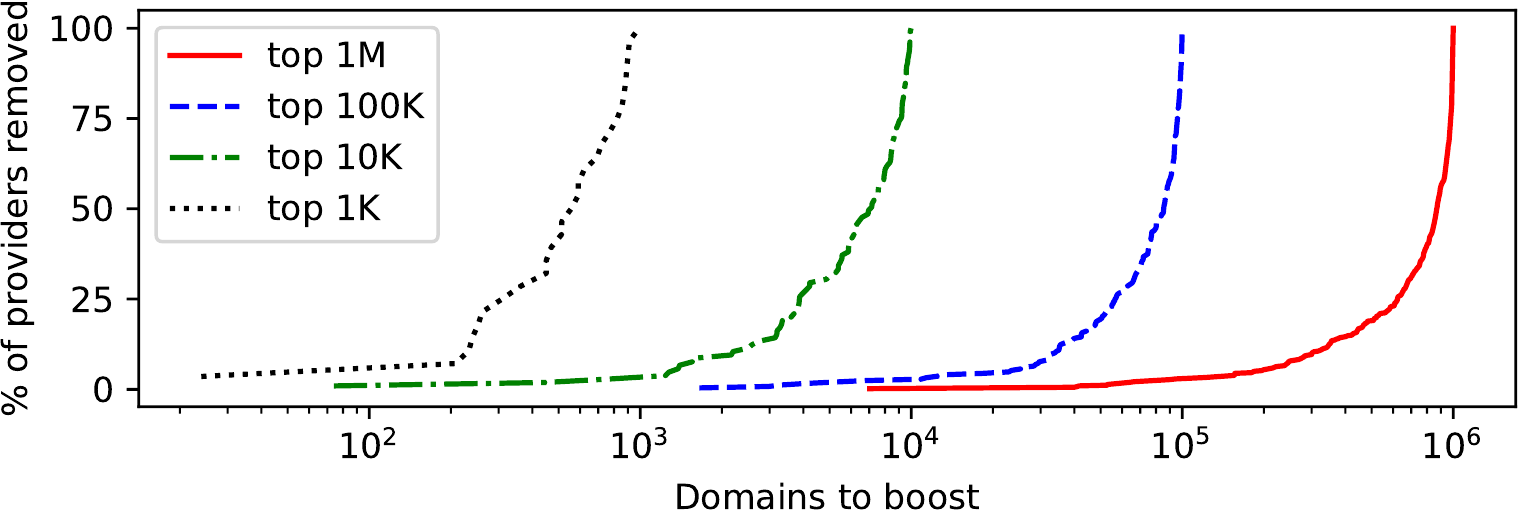}
	\caption{The percentage of fingerprinting script providers that would not be detected if a given number of domains were pushed above all fingerprinting domains for different subsets of Alexa's ranking.}
	\label{fig:required-manipulation}
\end{figure}

Figure~\ref{fig:required-manipulation} shows how many fingerprinting providers would fully disappear from the Alexa list if a given number of domains are manipulated. We consider removal for different subsets, as commonly used by the studies that we surveyed in Section~\ref{sub:survey}. The smallest number of manipulated domains required is 7\,032, 1\,652, 74 and 24 for the top 1M, 100K, 10K and 1K respectively; 15 providers need less than 100\,000 manipulated domains to disappear from the top 1M. 

As we will show, the cost of such large-scale manipulation is very low and well within reach of larger providers, especially given the incentive of being able to stealthily continue tracking. Moreover, this is an upper bound needed to remove all instances of a tracker domain from the list; reducing the prevalence of a script requires only hiding the worst-ranked domains. Finally, it is not required to insert new domains: forging a few requests to boost sites already in the list is sufficient, further reducing the cost and even making the manipulation harder to detect.

Englehardt and Narayanan highlighted how ``the long tail of fingerprinting scripts are largely unblocked by current privacy tools,'' reinforcing the potential impact of exposing these scripts. A malicious party can therefore gain an advantage by actively manipulating the rankings of popular domains. As we will show in the next section, such manipulation is actually feasible across all four lists, usually even on a sufficiently large scale without the need for significant resources.

\section{Feasibility of large-scale manipulation}\label{sec:manipulation}

The data collection processes of popularity rankings rely on a limited view of the Internet, either by focusing on one specific metric or because they obtain information from a small population. This implies that targeted small amounts of traffic can be deemed significant on the scale of the entire Internet and yield good rankings. Moreover, the ranking providers generally do not filter out automated or fake traffic, or domains that do not represent real websites, further reducing the share of domains with real traffic in their lists.

Consequently, attacks that exploit these limitations are especially effective at allowing arbitrary modifications of the rankings at a large scale. We showed how adversaries may have incentives to skew the conclusions of security studies, and that security researchers and practitioners often use popularity rankings to drive the evaluation of these studies. Manipulating these rankings therefore becomes a prime vector for influencing security research, and as we will show, the small costs and low technical requirements associated with this manipulation make this approach even more attractive.

For each of the four studied popularity rankings, we describe techniques that manipulate the data collection process through the injection of forged data. To prove their feasibility, we execute those techniques that conform to our ethical framework and that have a reasonable cost, and show which ranks can be achieved. In Table~\ref{tbl:manipulation-summary}, we summarize the techniques and the cost they incur on three aspects: money, effort and time required. Through this cost assessment, we identify how these manipulations could be applied at scale and affect a significant portion of these lists.

These techniques can be applied to both new domains and domains already present in the lists, e.g.\ when those domains bear the properties that could skew certain studies; a domain that has been ranked for a longer period of time may enjoy a higher trust or importance. In our work, we focus on techniques that directly influence the rankings' data at a modest cost. An alternative approach could be to buy expired or parked domains already in the list~\cite{Moura2017}. However, expired domains are usually bought up very quickly by ``drop-catchers''~\cite{Lever2016}, leaving a limited number of ranked domains open for registration~\cite{ExpiredDomains}. Meanwhile, popular parked domains can command prices upwards of 1\,000 USD~\cite{ExpiredDomains}. This approach therefore incurs a prohibitive cost, especially at a large scale. 

\begin{table}
	\centering
	\caption{Summary of manipulation techniques and their estimated cost.}
	\label{tbl:manipulation-summary}
	\begin{tabular}{llccc}
		\toprule
		& & \multicolumn{3}{c}{Cost}\\
		\cmidrule(lr){3-5}
		Provider & Technique & Monetary & Effort & Time \\
		\midrule
		Alexa & Extension & none & medium & low \\
		& Certify & medium & medium & high \\
		Umbrella& Cloud providers & low & medium & low \\
		Majestic & Backlinks & high & high & high \\
		& Reflected URLs & none & high & medium \\
		Quantcast & Quantified & low & medium & high \\ 
		\bottomrule
	\end{tabular}
\end{table}

\subsection{Alexa}\label{sub:manipulation-alexa}
Alexa ranks domains based on traffic data from two sources: their ``Traffic Rank'' browser extension that reports all page visits, and the ``Certify'' analytics service that uses a tracking script to count all visits on subscribing websites. We forge traffic data to both and observe the achieved ranks.

\subsubsection{Extension} 
The ``Alexa Traffic Rank'' extension collects data on all pages that its users visit. The extension also shows users information on the rank and traffic of the visited site, which may serve as an incentive to install the extension.

We submitted page visits for both registered and nonexistent test domains previously unseen by Alexa. We generated profiles with all 1\,152 possible configurations, i.e.\ the demographic details that are requested when installing the extension, and this within a short timeframe from the same IP address; Alexa did not impose any limits on the number of profiles that could be created. We submitted visits to one domain per profile; as visits to the same page by the same profile are only counted once~\cite{Alexa1}, we generated exactly one visit per page to the homepage and randomly generated subpages. The number of page views for one test domain ranges from 1 to 30.

We installed the extension in a real Chrome browser instance and then generated page visits to our test domain, simulating a realistic usage pattern by spacing out page visits between 30 and 45 seconds, and interspersing them with as many visits to domains in Alexa's top 1000. Through inspection of the extension's source code and traffic, we found that upon page load, a GET request with the full URL of the visited page\footnote{For pages loaded over HTTPS, the path is obfuscated.} is sent alongside the user's profile ID and browser properties to an endpoint on \url{data.alexa.com}. This means these requests can also be generated directly without the need to use an actual browser, greatly reducing the overhead in manipulating many domains on a large scale.

From May 10, 2018 onward, Alexa appears to block data reporting from countries in the European Union (EU) and European Economic Area (EEA), as the response changed from the visited site's rank data shown to the user to the string ``Okay''. This is likely due to the new General Data Protection Regulation coming into force. While we were able to circumvent this block through a VPN service, Alexa may be ignoring traffic in EU and EEA countries, introducing a further bias towards traffic from other countries.

For 20\% of our profiles/domains, we were successful in seeing our page views counted and obtaining rankings within the top million. Alexa indicates that it applies statistical processing to its data~\cite{Alexa2}, and we suspect that some of our requests and generated profiles were pruned or not considered sufficient to be ranked, either because of the profile's properties (e.g.\ a common browser configuration or an overrepresented demographic) or because only a subset of traffic data is (randomly) selected. To increase the probability of getting domains ranked, an adversary can select only the successful profiles, or generate page views to the same site with different profiles in parallel, improving the efficiency of their manipulation.

\begin{figure}
	\centering
	\subfloat[Extension.]{\includegraphics[width=0.45\columnwidth]{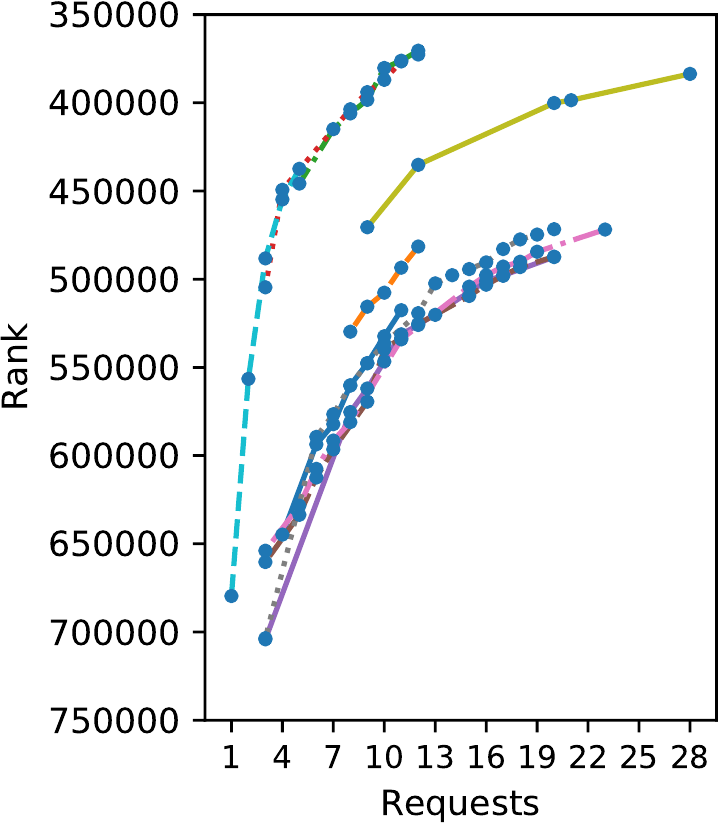}
		\label{fig:manipulation-alexa-extension}}
	\hfil
	\subfloat[Certify.]{\includegraphics[width=0.45\columnwidth]{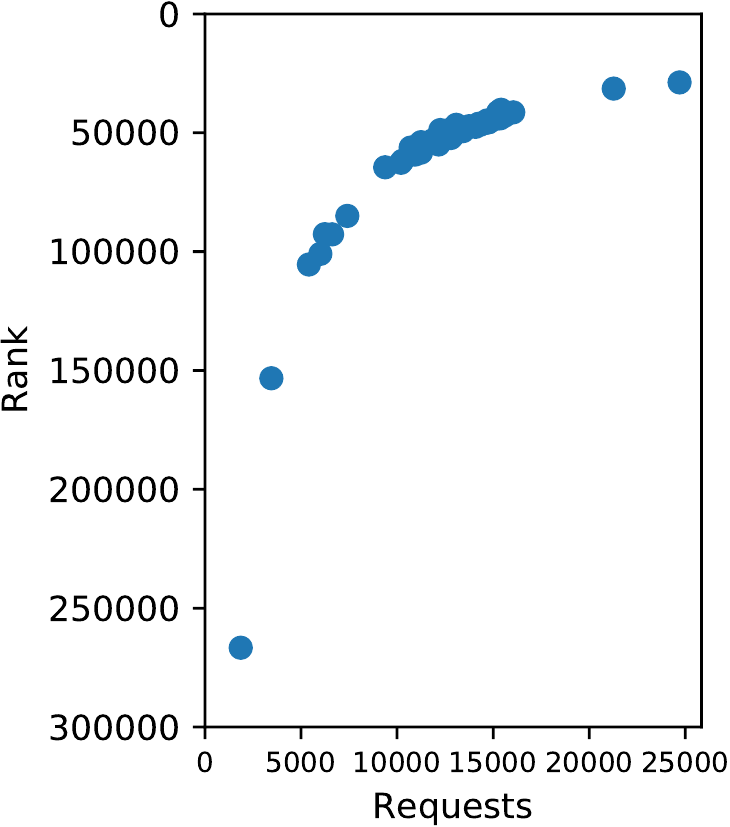}
		\label{fig:manipulation-alexa-certify}}
	\caption{Ranks obtained in the Alexa list. Ranks on the same day are connected.}
	\label{fig:manipulation-alexa}
\end{figure}

Figure~\ref{fig:manipulation-alexa}\subref{fig:manipulation-alexa-extension} lists our 224 successful rankings grouped per day, showing the relation between ranks and number of visits. We performed our experiments between July 25 and August 5, 2018. As during this period Alexa averaged traffic over one day, there was only a delay of one day between our requests and the domains being ranked; they disappeared again from the list the following day. This means that it is not necessary to forge requests over a longer period of time when the malicious campaign is short-lived.

What is most striking, is the very small number of page visits needed to obtain a ranking: as little as one request yielded a rank within the top million, and we achieved a rank as high as 370\,461 with 12 requests (albeit in the week-end, when the same number of requests yields a better rank).
This means that the cost to manipulate the rankings is minimal, allowing adversaries to arbitrarily alter the lists at large scale for an extended period of time. This ensures continued ranking and increases the likelihood of a list containing manipulated domains being used for research purposes, despite the large daily change. 

The low number of required requests is further confirmed by large blocks of alphabetically ordered domains appearing in the ranking: these point towards the same number of visits being counted for these domains. We use these blocks as well as the processed visitor and view metrics retrieved from the Alexa Web Information Service~\cite{Alexa9} to estimate the required visit count for better ranks.

\begin{figure}
	\centering
	\includegraphics[width=\columnwidth]{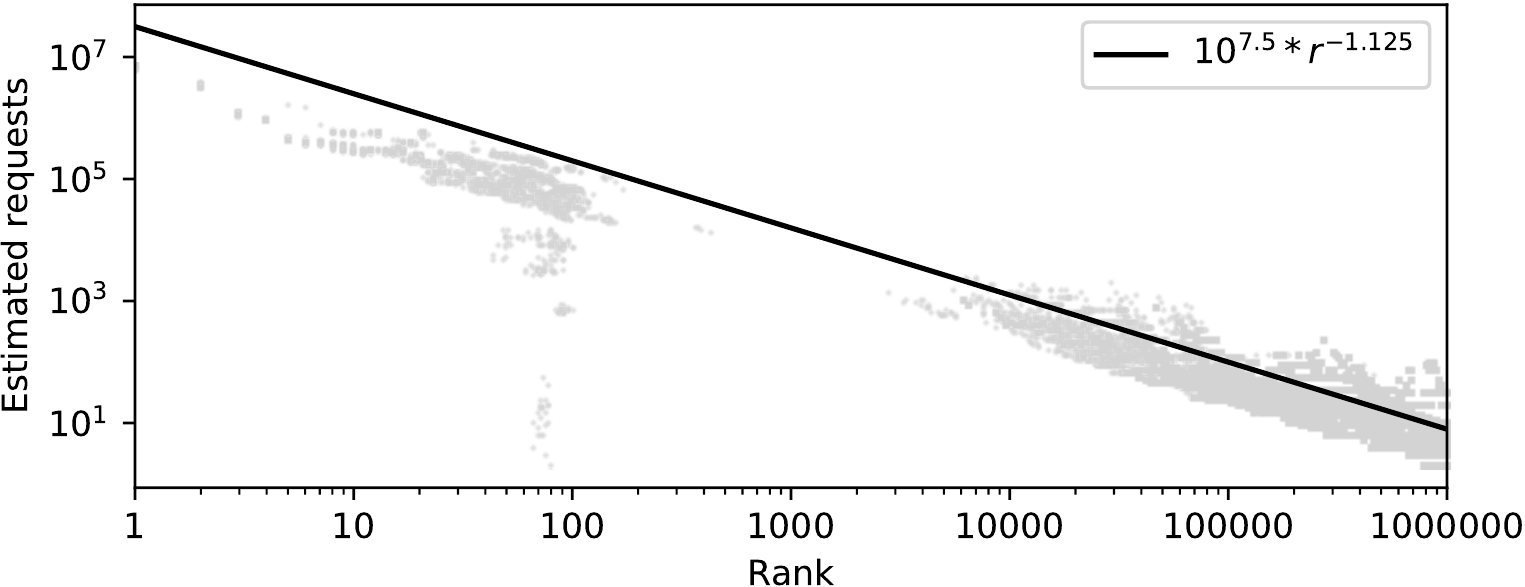}
	\caption{The estimated relation between requests and rank for Alexa. The gray areas show data as retrieved from the Alexa Web Information Service.}
	\label{fig:alexa-estimate}
\end{figure}

Figure~\ref{fig:alexa-estimate} shows the number of requests needed to achieve a certain rank; we consider this an upper bound as Alexa ranks domains that see more unique visitors better than those with more page views, meaning that manipulation with multiple profiles would require less requests. This analysis shows that even for very good ranks, the amount of requests required and accompanying cost remains low, e.g.\ only requiring 1\,000 page views for rank 10\,000. This model of Alexa's page visits also corresponds with previous observations of Zipf's law in web traffic~\cite{Adamic2002a, Clauset2009}.

Alexa's list is also susceptible to injection of nonexistent domains; we were able to enter one such domain. Furthermore, we confirmed in our server logs that none of our test domains were checked by Alexa as we forged page visit requests. The ability to use fake domains reduces the cost to manipulate the list at scale even further: an attacker is not required to actually purchase domain names and set up websites for them. 

Even though Alexa's statistical postprocessing may prune some visits, the low number of required visits, the ability to quickly generate new profiles and the lack of filtering of fake domains allows an attacker to still easily achieve significant manipulation of Alexa's list.

\subsubsection{Certify}
Alexa's `Certify' service offers site owners an analytics platform, using a tracking script installed on the website to directly measure traffic. The service requires a subscription to Alexa's services, which start at USD 19.99 per month for one website.

As Alexa verifies installation of its scripts before tracking visits, we installed them on a test website. From the JavaScript part of this code, we extracted its reporting algorithm and repeatedly forged GET requests that made us appear as a new user visiting the website, therefore avoiding the need to retain the response cookies for continued tracking. To diversify the set of IP addresses sending this forged traffic, we sent these requests over the Tor network, which has a pool of around 1\,000 IP addresses~\cite{TorRelays}. We sent at most 16\,000 requests per 24 hours, of which half were for the root page of our domain, and the other half for a randomly generated path. 

Figure~\ref{fig:manipulation-alexa}\subref{fig:manipulation-alexa-certify} lists the ranks of our test domain and the number of visits that were logged by Alexa across 52 days. For 48 days, we reached the top 100\,000 (purported to more accurately reflect popularity), getting up to rank 28\,798.
Not all our requests were seen by Alexa, but we suspect this is rather due to our setup (e.g.\ by timeouts incurred while sending requests over Tor). Alexa's metrics report that our site received "100.0\% real traffic" and that no traffic was excluded, so we suspect that Alexa was not able to detect the automated nature of our requests. 

After subscription to the service, Alexa will only calculate (and offer to display) the `Certified' rank of a website after 21 days. Since no visits to our site were being reported through Alexa's extension, no `normal' rank was achieved in the meantime, and therefore there was a large delay between the start of the manipulation and the ranking of the domain.

The disadvantage of this technique is that the cost of manipulation at scale quickly becomes prohibitive, as for each site that needs to be inserted into the list, a separate subscription is required. Given Alexa's verification of the tracking script being installed, the domain needs to be registered and a real website needs to be set up, further reducing the scalability of the technique. However, we were able to achieve better ranks with a more consistent acceptance of our forged requests. Depending on the attacker's goal, it is of course still possible to artificially increase the ranking of specific websites who already purchased and installed the Alexa Certify service.

We obtained a rank even though we did not simulate traffic to this test domain through the Alexa extension, which strongly suggests that Alexa does not verify whether `Certified' domains show similar (credible) traffic in both data sources.
Based on this observation, we found one top~100 `Certified' site where Alexa reports its extension recording barely any or even no traffic: while in this case it is a side-effect of its usage pattern (predominantly mobile), it implies that manipulation conducted solely through the tracking script is feasible.

\subsection{Cisco Umbrella}
Umbrella ranks websites on the number of unique client IPs issuing DNS requests for them. Obtaining a rank therefore involves getting access to a large variety of IP addresses and sending (at least) one DNS request from those IPs to the two open DNS resolvers provided by Umbrella.

\subsubsection{Cloud providers}
Cloud providers have obtained large pools of IP addresses for distribution across their server instances; e.g.\ Amazon Web Services (AWS) owns over 64 million IPv4 addresses~\cite{AWS1}. These can be used to procure the unique IP addresses required for performing DNS requests, but due to their scarcity, providers restrict access to IPv4 addresses either in number or by introducing a cost. 

In the case of AWS, there are two options for rapidly obtaining new IPv4 addresses. Continuously starting and stopping instances is an economical method, as even 10\,000 different IPs can be obtained for less than USD 1 (using the cheapest instance type), but the overhead of relaunching instances reduces throughput: on the cheapest \texttt{t2.nano} instance, we were able to obtain a new IP on average every minute. Moreover, the number of concurrent running instances is limited, but by using instances in multiple regions or even multiple accounts, more instances are accessible.  
Keeping one instance and allocating and deallocating Elastic IP addresses (i.e.\ addresses permanently assigned to a user) yields higher throughput, at 10 seconds per IP. However, AWS and other providers such as Microsoft Azure discourage this practice by attaching a cost to this `remap' operation: for AWS, a remap costs USD 0.10, so a set of 10\,000 IPs incurs a prohibitive cost of USD 1\,000.

\begin{figure}
	\centering
	\includegraphics[width=\columnwidth]{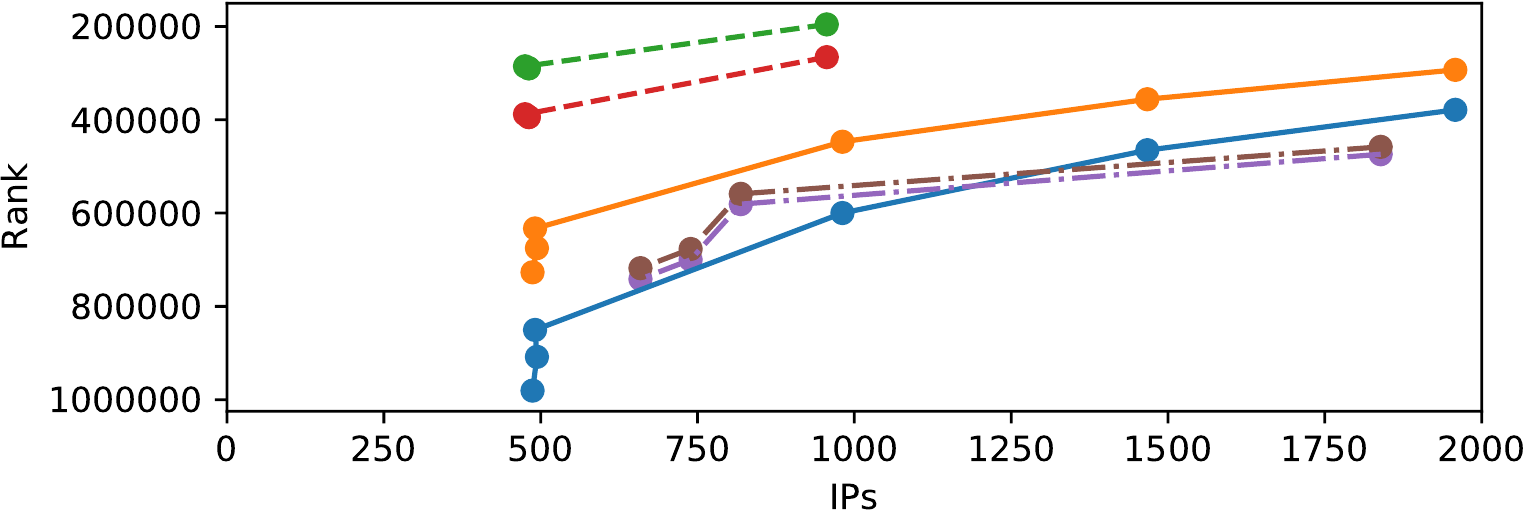}
	\caption{Ranks obtained in the Umbrella list. Ranks on the same day are connected; ranks over two days for one set of requests use the same line style.}
	\label{fig:manipulation-umbrella}
\end{figure}

Figure~\ref{fig:manipulation-umbrella} shows the relation between the number of issued DNS requests and the obtained rank; all of our attempts were successful. We were able to obtain ranks as high as 200\,000 with only a thousand unique IP addresses, albeit in the weekend, when OpenDNS processes around 30\% less DNS traffic~\cite{OpenDNSSystem}. We only sustained DNS traffic for one day at a time, but it appears that Umbrella counts this traffic (and therefore ranks the domain) for two days, reducing the number of requests needed per day to either obtain a good rank for one domain or rank many domains.

Given the relatively high cost per IP, inserting multiple domains actually is more economical as several DNS requests can be sent for each IP instantiation. As the name requested in the DNS query can be chosen freely, inserting fake domains is also possible; the high number of invalid entries already present shows that Umbrella does not apply any filtering. This further improves scalability of this technique, as no real websites need to be set up in order to manipulate the list.

The effort to generate many ranked entries is further reduced by the inclusion of subdomains, as all subdomains at lower depths are automatically ranked: we were able to rank 12 subdomains simultaneously with one set of requests. Furthermore, the number of requests is aggregated per subdomain, so a low number of requests to many subdomains can result in both many ranked subdomains and a good rank for the pay-level domain.

Combining the ability to insert fake domains with the low overhead of requests to additional domains, the inclusion of subdomains and the lack of any filtering or manipulation detection means that the scale at which an attacker can manipulate Umbrella's list can be very large.

\subsubsection{Alternatives}
\begin{itemize}[wide]
	\item \textit{Tor.} The Tor service provides anonymous communication between a user and the service they use. Traffic is relayed across multiple nodes before being sent to the destination from an exit node, meaning that the destination observes traffic originating from that node's IP address. This set of exit nodes provide a pool of IP addresses, and by switching the routing over the Tor network, DNS requests can be altered to appear to originate from multiple IP addresses in this pool. However, as there are less than 1\,000 exit nodes at any given point in time~\cite{TorRelays}, it will be possible to inject domains in the list, but infeasible to obtain a high rank solely through this technique.
	
	\item \textit{IP spoofing.} IP packets contain the IP address of its sender, that can however be arbitrarily set in a technique known as IP spoofing. We could leverage this technique to set the source IP of our DNS packets to many different addresses, in order for our requests to appear for Umbrella to originate from many unique IPs. As IP spoofing is often used during denial-of-service attacks, many ISPs block outgoing packets with source IPs outside their network. Leveraging IP spoofing for sending DNS requests therefore requires finding a network that supports it. Karami \etal~\cite{Karami2016} found that certain VPS providers allow IP spoofing; as such these could be used for our experiment.
	
	Due to the ethical concerns that are raised by leveraging IP spoofing (the responses of our DNS requests would arrive at the users of the forged source IPs, and the associated traffic may cause the VPS provider to be flagged as malicious), we did not further explore this technique. It is important to note however that an adversary only needs to find a single provider or network that does not prevent IP spoofing in order to send a very large number of DNS requests to Umbrella's resolvers and thus manipulate the list at a very large scale.
\end{itemize}

\subsection{Majestic}
Majestic's ranking is based on the number of subnets hosting a website that links to the ranked domain. Therefore, we cannot construct data reporting requests sent directly to Majestic, but must use techniques where website owners knowingly or unknowingly serve a page that contains a link to our domain and that is then crawled independently by Majestic. 

\subsubsection{Backlinks}
Backlink providers offer a paid service where they place incoming links for a requested website (`backlinks') on various sites. The goal of this service is usually to achieve a higher position in search engine rankings, as part of search engine optimization (SEO) strategies; the deceptive nature of this technique makes that this is considered `black-hat' SEO.

Backlinks are priced differently according to the reputation of the linking site. While we need a sufficiently diverse set of websites hosted on different subnets, Majestic does not take the quality of our backlinks into account when ranking domains. This means that we can reduce our cost by choosing the cheapest type of backlink.
Moreover, we have the choice of removing backlinks after they have been found, as these are no longer billed but still count towards the subnets for a period of at most 120 days, reducing monetary cost.

We use the services of \texttt{BackLinks.com}, as they operate only on sites under their control, therefore avoiding impact of our experiment on unaware site owners. The choice for this particular backlink provider brings about certain constraints (such as the pool of available backlink sites, or a limit on daily backlink deletions), but these can be alleviated by using other and/or multiple backlink providers. We buy backlinks if they are located in a subnet not covered by any already purchased site, but have to use OCR as the URLs on which links would be placed are only available as a warped image. We therefore curated the set of backlinks through manual verification to compensate for any errors, increasing our required effort.

The cheapest type of backlink costs USD 0.25 a month, but since there was not a sufficient amount of such pages to cover the necessary number of subnets, more expensive backlinks were also required. The backlinks were partially found organically by Majestic; in this case there is no additional cost. Through a subscription on Majestic's services, backlinks can also be submitted explicitly for crawling: the minimum cost is USD 49.99 for one month.

\begin{figure}
	\centering
	\includegraphics[width=\columnwidth]{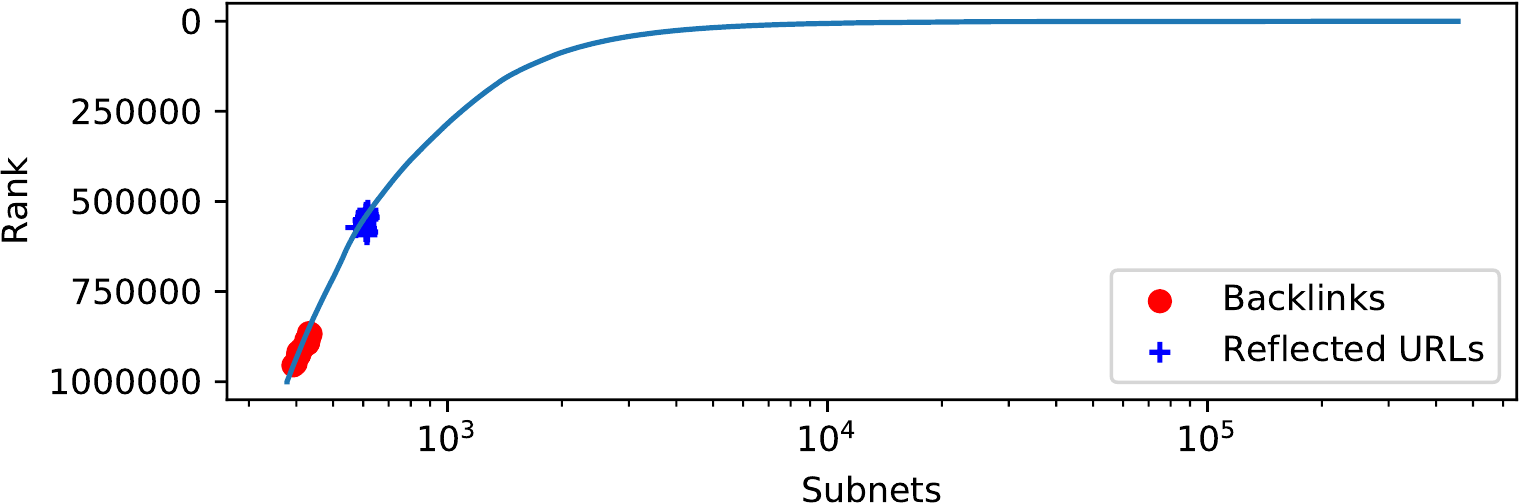}
	\caption{The relation between subnets and rank in the Majestic list for May 31, 2018, with our obtained ranks highlighted.}
	\label{fig:manipulation-majestic}
\end{figure}

We bought backlinks for our test domain and curated them for two and a half months, in order to capture as many subnets as possible while managing the monetary cost. Our total cost was USD 500. We successfully inserted our domain, with Figure~\ref{fig:manipulation-majestic} showing the achieved rankings on top of the relation between the rank and the number of found subnets for all ranked sites as published by Majestic.

There exists a trade-off between the cost and the time required to enter the rank: if the monetary cost should be kept low, more time is needed as the set of eligible backlink pages is smaller and backlinks will need to be deleted. Alternatively, a higher number of possibly more expensive backlinks would allow to achieve the necessary number of subnets more quickly, but at a higher monetary cost. Conversely, because Majestic considers links for at least 120 days, the cost for long-term manipulation is relatively limited: even though we stopped buying backlinks and these subsequently disappeared, our ranking was still maintained for more than two months as previously found backlinks were still counted.

\subsubsection{Reflected URLs}
An alternative technique that we discovered, for which it is not required to purchase services from external parties, is to leverage websites that reflect a GET parameter into a link. Note that for our purpose, reflected cross-site scripting (XSS) attacks could also be used; however, this technique is more intrusive as it will inject HTML elements, so we did not evaluate it out of ethical considerations.
To discover web pages that reflect a URL passed as a parameter, we started crawling the 2.8 million domains from the four lists, finding additional pages by following links from the homepage of these domains. If GET parameters were found on the page, we replaced each one with a URL and tested whether this URL was then included in the \texttt{href} of an \texttt{a} tag on the page.

Through this crawl, we found that certain MediaWiki sites were particularly susceptible to reflecting URLs on each page, depending on the configuration of the site. We therefore tested this reflection on the wikis from a number of data sources: the root domains as well as the subdomains containing \texttt{wiki} of the four top lists, the set of wikis found by Pavlo and Shi in 2011~\cite{Pavlo2011} and the wikis found by WikiTeam\footnote{\url{https://github.com/WikiTeam/wikiteam}}. As the reflection is purely achieved through altering the GET parameters, we do not permanently alter the wiki.

Given the special construction of their URLs, the pages reflecting our domain will not be found organically by Majestic. The list of affected URLs can be submitted directly to Majestic, but this requires a subscription. The links can also be placed on one aggregating web page: by verifying ownership of the hosting domain with Majestic, a crawl of this page and subsequently of the links placed on it can be triggered for free; alternatively, using Majestic's site to request the freely available subset of backlinks data for this special web page also seems to trigger this crawl.

Through our crawls, we found 1\,041 pages that reflected the URL of our test domain when passed in a GET parameter. Through submitting these reflecting URLs to Majestic's crawler, we successfully ranked our domain, with Figure~\ref{fig:manipulation-majestic} showing the achieved rankings over time. Through this technique, we also successfully had one backlink to a non-existing domain crawled and counted as a referring subnet. By scaling this up to the number of subnets required to be ranked, this implies that Majestic's list ranking is also susceptible to fake entries; as there are unavailable sites in the list, Majestic likely does not actively check whether entries in the list are real.

This technique allows to construct backlinks at no monetary cost, but requires a high effort to find appropriate pages. We found only small subsets of wikis and domains in general to reflect our URL, so the number of pages and subnets that can be discovered using this technique may not be sufficient to achieve very high rankings. Given a deeper crawl of pages, more sites that reflect URLs passed through a GET parameters may be found, more subnets can be covered and a higher ranking can be achieved. Moreover, an attacker can resort to more `aggressive' techniques where URLs are permanently stored on pages or XSS vulnerabilities are exploited.

Once found however, a reflecting URL will be counted indefinitely: a site would effectively have to be reconfigured or taken offline in order for the backlink to disappear. This means maintaining a rank comes at no additional cost. Furthermore, every website that is susceptible to URL reflection can be leveraged to promote any number of attacker-chosen (fake) domains, at the cost of submitting more (crafted) URLs to Majestic. This means that manipulation of Majestic's list is also possible on a large scale. 

\subsubsection{Alternatives}
\begin{itemize}[wide]
	\item \textit{Hosting own sites.} Using domains seen in passive DNS measurements, Tajalizadehkhoob \etal~\cite{Tajalizadehkhoob2016} identified 45\,434 hosting providers in 2016, and determined their median address space to contain 1\,517 IP addresses. Based on these figures, we can assume that the number of subnets available through hosting providers is well above the threshold to be ranked by Majestic. An attacker could therefore set up websites on a sufficient number of these providers, all with a link back to the domain to be ranked. By making all the websites link to each other, a larger set of domains could easily be ranked. This technique incurs a high cost however: in effort, as setting up accounts with these providers is very likely to require a lot of manual effort, as well as in monetary cost, as for each hosting provider a subscription needs to be bought.
	\item \textit{Pingbacks.} Content management systems such as WordPress provide a pingback mechanism for automatically reporting URLs that link to one of the pages hosted on that system. Many sites will then insert a link back to the reported URL on that page. By finding a set of domains supporting pingbacks (similar to finding wikis) and reporting a URL on the domain we want to see ranked, we could again have links to our domain on a large set of domains and therefore subnets. However, this permanently changes pages on other websites, and although enabling the pingback feature implies some consent, we opted to not explore this technique for ethical reasons.
\end{itemize}

\subsection{Quantcast}
\subsubsection{Quantified}
Quantcast mainly obtains traffic data through its tracking script that webmasters install on their website. We extracted the reporting algorithm from the tracking script, and automatically sent requests to Quantcast from a set of 479~VPN servers located in the United States, as Quantcast's ranking only takes US traffic into account. We sent requests for 400 generated users per day, presenting ourselves as a new user on the first request and subsequently reusing the generated token and received cookie in four more requests. As opposed to Alexa's tracking script, reporting page views for only new users did not result in any visits being counted. 

Our forged requests were acknowledged by Quantcast and its analytics dashboard reports that on May 30, 2018, "the destination reaches over 6,697 people, of which 6,696 (100\%) are in the U.S." The latter metric is used to determine the rank. However, our test domain has not appeared in the ranking. This is likely due to the short age of our domain; although we have sent requests for more than a month, Quantcast's slow update frequency means its ranking algorithm may not take our domain into account yet.

\begin{figure}
	\centering
	\includegraphics[width=\columnwidth]{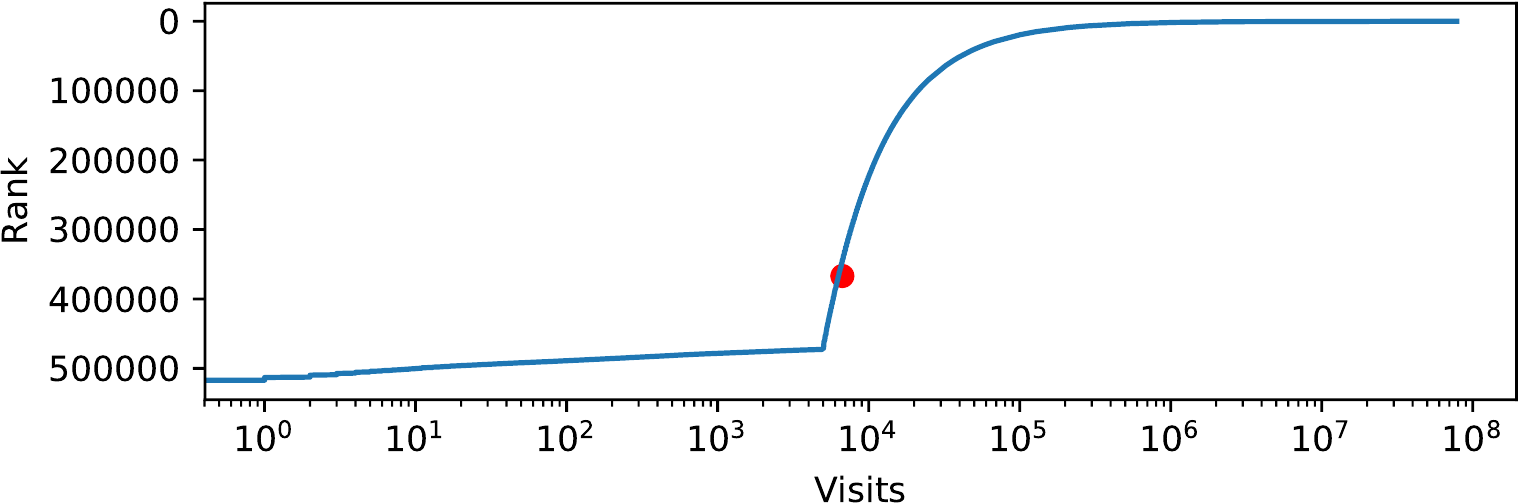}
	\caption{The relation between measured visits and rank in the Quantcast list for May 31, 2018, with the theoretical rank for our visit count highlighted.}
	\label{fig:manipulation-quantcast}
\end{figure}

As Quantcast publishes the number of visits counted for each ranked domain, the relation between the desired rank and required effort is known as shown in Figure~\ref{fig:manipulation-quantcast}. Up to around 5\,000 visits, the achieved rank remains relatively low; this tail contains primarily quantified sites that are ranked even with almost no visits. Above 5\,000 visits, Quantcast's list includes many more domains for which a rank is estimated; especially at worse ranks, large blocks of estimated domains are interspersed with quantified domains, so increasing the number of visits to jump across such a block gives a large improvement in rank. If a rank were to be assigned to our domain, we can determine that we would theoretically be given a rank around 367\,000. Achieving higher ranks only requires submitting more forged requests, so the increased cost in time and effort is minimal.

Quantcast will only start processing traffic data once it has verified (through crawling) that its tracking pixel is present on the domain. It is therefore required to register the domain and set up a real website to manipulate the rankings, so scaling to multiple domains incurs a higher cost; Quantcast's analytics platform itself is free however, limiting the additional cost. As Quantcast performs the check only once, the domain and the website also do not need to be sustained. Merely registering for tracking may even suffice to be ranked: over 2\,000 domains are ranked but reported to have 0~visits, with over half registered by DirectEmployers as discussed in Section~\ref{sub:representativeness}.

\subsubsection{Alternatives}
Quantcast states that it also uses traffic data from `ISPs and toolbar providers'~\cite{Quantcast1}. ISPs sell traffic data to third parties~\cite{ISPSellData}, and Quantcast may be buying these services to generate the number of page visits and therefore the rank for non-quantified websites. However, we cannot determine which ISPs may be used. As for extensions, we were unable to discover any extensions reporting to a URL that was obviously related to Quantcast.

\paragraph*{\textbf{Ethical considerations}} Because our experiments may have a large impact on the reputation of the rankings as well as potentially affect third parties, we conduct an ethical review of our experimental methods. Such reviews have been advocated for by the academic community~\cite{Partridge2016} and ensure that the potential damage inflicted is minimized. We base this review on the ethical principles outlined in the Menlo Report~\cite{Dittrich2012}, which serves as a guideline within the field of ICT research; we apply the principle of beneficence in particular: identifying potential benefits and harms, weighing them against each other and minimizing the risk of inflicting harm.

Because of their commercial nature, the providers of popularity rankings have an economic interest in these being accurate. We show that these lists can be manipulated, negatively affecting their perceived reputability. Our findings are however of value to the providers: by evaluating the various techniques and reporting our findings, the providers become aware of the potential threats, may take actions to thwart attackers and can improve the correctness of their rankings. 

We have disclosed our findings and proposals for potential remedies to the four providers, alongside a list of manipulated domains for them to remove from their datasets and past and present rankings. Alexa and Majestic provided statements regarding the value of their rankings and the (in)feasibility of manipulation, but commercial considerations prevent them from elaborating on their methods. Cisco Umbrella closed our issue without any statement, and we received no response from Quantcast. None of our test domains were (retroactively) removed from any rankings after our notification.

We minimize the impact of our experiments on third parties by only significantly manipulating the ranking of our own, purposefully registered domains and refraining from intrusive or questionable techniques. Our sites also contained an explanation of our experiment and contact details for affected parties. Our low number of test domains means that only few domains will see negligible shifts in ranking due to our experiments; e.g.\ the volatility of Alexa's list has a significantly larger impact. Moreover, we minimized the duration of our experiments and of our domains being ranked. The impact on other research using these lists is also minimal; we showed that in general many more ranked domains are unavailable or unrepresentative. Our sites only hosted benign content, so whitelists using rankings are unaffected.

\section{An improved top websites ranking}

As we showed, the different methods used to generate popularity rankings cause undesirable effects on their properties that can potentially sway the results and conclusions of studies. In addition, we showed that researchers are prone to ignore or be unaware of these effects. We also proved that these rankings show several pitfalls that leave them vulnerable to large-scale manipulation, further reducing their reliability and suitability to research. Nevertheless, popularity rankings remain essential for large-scale empirical evaluations, so we propose improvements to existing rankings as well as a new ranking that has characteristics geared towards research.

\subsection{Defending existing rankings against manipulation}
Even though the methods for data collection and processing of the existing lists are usually unknown, our experiments suggest that their providers employ little defense against large-scale manipulation. We outline techniques that the providers could use to make these lists more resilient to attacks.

Detecting and deterring singular instances of fraud ensures that all data used in ranking domains is deemed valid. Alexa and Quantcast rely on the reporting of page visits; within the realm of online advertising, techniques have been designed to subvert click inflation~\cite{Blundo2010,AbuRajab2008a,Metwally2007}.
As we saw that not all attempts at manipulating Alexa's ranking were successful, this may imply that Alexa already employs some of these tactics.

To deter large-scale manipulation, ranking providers could employ tactics that increase the effort and resources required to affect many domains to prohibitive levels. This therefore avoids significant influence on research results, even if these tactics may not be sufficient to stop small-scale manipulation. 

For a traffic reporting extension, the profile setup could be tied to an account at an online service; while a normal user can easily create one account, creating many accounts in an automated way can be countered by techniques that try to detect fake accounts~\cite{Cao2012}. In the case of Alexa, given its ownership by Amazon, a natural choice would be to require an Amazon account; in fact, a field for such an account ID is available when registering the extension, but is not required. This technique is not useful for tracking scripts, since no user interaction can be requested, and fraud detection as discussed earlier may be required. For providers that use both, the two metrics can be compared to detect anomalies where only one source reports significant traffic numbers, as we suspect such manipulation is already happening for Alexa Certify.

Data could be filtered on the IP address from which it originates. Ignoring requests from ranges belonging to cloud providers or conversely requiring requests to come from ranges known to belong to Internet service providers (e.g.\ through its autonomous system) does not block a single user from reporting their traffic. However, using many IP addresses concurrently is prevented as these cannot be easily obtained within the permitted ranges. This technique is particularly useful for Umbrella's list; for the other lists, using many IP addresses is not strictly necessary for large-scale manipulation.

The relative difficulty of maliciously inserting links into pages on many IP subnets already reduces the vulnerability of link-based rankings to large-scale manipulation. Specific attacks where the page reflects a URL passed as a parameter could be detected, although this can be made more difficult by obfuscation and attacks that alter a page more permanently. The link-based rankings could be refined with reputation scores, e.g. the age of a linked page or Majestic's ``Flow Metrics''~\cite{Majestic1}, to devalue domains that are likely to be part of a manipulation campaign.

Finally, requiring ranked domains to be available and to host real content increases the cost of large-scale manipulation, as domain names need to be bought and servers and web pages need to be set up. For Umbrella, not ranking domains where name resolution fails can significantly reduce unavailable (and therefore possibly fake) domains in the list. The other providers can perform similar availability checks in the DNS or by crawling the domain.

\subsection{Creating rankings suitable for research}

As we cannot ensure that providers will (want to) implement changes that discourage (large-scale) manipulation, we look at combining all currently available ranking data with the goal of improving the properties of popularity rankings for research, canceling out the respective deficiencies of the existing rankings. To this extent, we introduce \listname, a service that researchers can use to obtain lists with such more desirable and appropriate properties. We provide standard lists that can be readily used in research, but also allow these lists to be highly configurable, as depending on the use case, different traffic sources or varying degrees of stability may be beneficial. 

Moreover, we provide a permanent record to these new lists, their configuration and their construction methods. This makes historical lists more easily accessible to reduce the effort in replicating studies based upon them, and ensures that researchers can be aware of the influences on the resulting list by its component lists and configuration.

Our service is available at \textbf{\url{https://tranco-list.eu}}. The source code is also openly published at \url{https://github.com/DistriNet/tranco-list} to provide full transparency of how our lists are processed.

\subsubsection{Combination options and filters}
We support creating new lists where the ranks are averaged across a chosen period of time and set of providers, and introduce additional filters, with the goal of enhancing the research-oriented properties of our new lists. 

In order to improve the rank of the domains that the lists agree upon, we allow to average ranks over the lists of some or all providers.  We provide two combination methods: the Borda count where, for a list of length $N$, items are scored with $N, N-1, ..., 1, 0$ points; and the Dowdall rule where items are scored with $1, 1/2, ..., 1/(N-1), 1/N$ points~\cite{Fraenkel2014}. The latter reflects the Zipf's law distribution that website traffic has been modeled on~\cite{Adamic2002a, Clauset2009}. Our standard list applies the Dowdall rule to all four lists. We also allow to filter out domains that appear only on one or a few lists, to avoid domains that are only marked as popular by one provider: these may point to isolated manipulation.

To improve the stability of our combined lists, we allow to average ranks over the lists of several days; our standard list uses the lists of the past 30~days. Again, we allow to filter out domains that appear only for one or a few days, to avoid briefly popular (or manipulated) domains. Conversely, if capturing these short-term effects is desired, lists based on one day's data are available. When combining lists, we also provide the option to only consider a certain subset of the input lists, to select domains that are more likely to actually be popular.

Differences in list composition complicate the combination of the lists. Umbrella's list includes subdomains; we include an option to use a recalculated ranking that only includes pay-level domains. Quantcast's list contains less than one million domains; we proportionally rescale the scores used in the two combination methods to the same range as the other lists.

We add filters to create a list that represents a certain desired subset of popular domains. A researcher can either only keep domains with certain TLDs to select sites more likely to be associated with particular countries or sectors, or exclude (overly represented) TLDs. To avoid the dominance of particular organizations in the list, a filter can be applied where only one domain is ranked for each set of pay-level domains that differ only in TLD. Finally, only certain subdomains can be retained, e.g.\ to heuristically obtain a list of authentication services by selecting \texttt{login.*} subdomains.

To allow researchers to work with a set of domains that is actually reachable and representative of real websites, we provide options to filter the domains on their responsiveness, status code and content length. We base these filters on a regular crawl of the union of all domains on the four existing lists. This ensures that the sample of domains used in a study yields results that accurately reflect practices on the web.

To further refine on real and popular websites, we include a filter on the set of around 3 million distinct domains in Google's Chrome User Experience Report, said to be `popular destinations on the web'~\cite{CrUX1}. Its userbase can be expected to be (much) larger than e.g.\ Alexa's panel; however, Google themselves indicate that it may not fully represent the broader Chrome userbase~\cite{CrUX1}. Moreover, the list is only updated monthly and does not rank the domains, so it cannot be used as a replacement for the existing rankings. 

To reduce the potential effects of malicious domains on research results (e.g.\ in classifier accuracy), we allow to remove domains on the Google Safe Browsing list~\cite{SafeBrowsing} from our generated lists.

\subsubsection{Evaluation}
We evaluate the standard options chosen for our combined lists on their improvements to similarity and stability; the representativeness, responsiveness and benignness of the included domains can be improved by applying the appropriate filters. We generate our combined lists from March 1, 2018 to November 14, 2018, to avoid distortions due to Alexa's and Quantcast's method changes, and truncate them to one million domains, as this is the standard for current lists.

\paragraph{Similarity} To determine the weight of the four existing lists, we calculate the rank-biased overlap with our combined lists. Across different weightings, the RBO with Alexa's and Majestic's lists is highest at 46.5--53.5\% and 46.5--52\% respectively, while the RBO with Quantcast's and Umbrella's lists is 31.5--40\% and 33.5--40.5\% respectively. These results are affected by the differences in list composition: subdomains for Umbrella and the shorter list for Quantcast mean that these two lists have less entries potentially in common with Alexa and Majestic, reducing their weight. Overall, there is no list with a disproportionate influence on the combined list.

\paragraph{Stability} Averaging the rankings over 30 days is beneficial for stability: for the list combining all four providers, on average less than 0.6\% changes daily, even for smaller subsets. For the volatile Alexa and Umbrella lists, the improvement is even more profound: the daily change is reduced to 1.8\% and 0.65\% respectively. This means that the data from these providers can be used even in longitudinal settings, as the set of domains does not change significantly.

\subsubsection{Reproducibility}
Studies rarely mention the date on which a ranking was retrieved, when the websites on that list were visited and whether they were reachable. Moreover, it is hard to obtain the list of a previous date: only Cisco Umbrella maintains a public archive of historical lists~\cite{Umbrella1}. These two aspects negatively affect the reproducibility of studies, as the exact composition of a list cannot be retrieved afterwards.

In order to enhance the reproducibility of studies that use one of our lists, we include several features that are designed to create a permanent record that can easily be referenced. Once a list has been created, a permanent short link and a preformatted citation template are generated for inclusion in a paper. Alongside the ability to download the exact set of domains that the list comprises, the page available through this link provides a detailed overview of the configuration used to create that particular list and of the methods of the existing rankings, such that the potential influences of the selected method can be assessed. This increases the probability that researchers use the rankings in a more well founded manner.

\subsubsection{Manipulation}

Given that our combined lists still rely on the data from the four existing lists, they remain susceptible to manipulation. As domains that appear on all lists simultaneously are favored, successful insertion in all lists at once will yield an artificially inflated rank in our combined list.

However, the additional combinations and filters that we propose increase the effort required to have manipulated domains appear in our combined lists. Averaging ranks over a longer period of time means that manipulation of the lists needs to be maintained for a longer time; it also takes longer for the manipulated domains to obtain a (significant) aggregated rank. Moreover, intelligently applying filters can further reduce the impact of manipulation: e.g.\ removing unavailable domains thwarts the ability to use fake domains. 

As each ranking provider has their own traffic data source, the effects of manipulating one list are isolated. As none of the lists have a particularly high influence in the combined list, all four lists need to manipulated to the same extent to achieve a comparable ranking in the combined list, quadrupling the required effort. For the combined list generated for October 31, 2018, achieving a rank within the top million would require boosting a domain in one list to at least rank 11\,091 for one day or rank 332\,778 for 30 days; for a rank within the top 100\,000, ranks 982 and 29\,479 would be necessary respectively. This shows that massive or prolonged manipulation is required to appear in our combined list.

\section{Related work}

The work that is most recent and most closely related to ours is that of Scheitle \etal~\cite{Scheitle2018a}, who compared Alexa's, Majestic's and Umbrella's lists on their structure and stability over time, discussed their usage in (Internet measurement) research through a survey of recent studies, calculated the potential impact on their results, and drafted guidelines for using the rankings. We focus on the implications of these lists for security research, expanding the analysis to include representativeness, responsiveness and benignness. Moreover, we are the first to empirically demonstrate the possibility of malicious large-scale manipulation, and propose a concrete solution to these shortcomings by providing researchers with improved and publicly available rankings.

In 2006, Lo and Sedhain~\cite{Lo2006} studied the reliability of website rankings in terms of agreement, from the standpoint of advertisers and consumers looking for the most relevant sites. They discussed three ranking methods (traffic data, incoming links and opinion polls) and analyzed the top 100 websites for six providers, all of which are still online but, except for Alexa, have since stopped updating their rankings.

Meusel \etal~\cite{Meusel2015} published one-time rankings of websites\footnote{\url{http://wwwranking.webdatacommons.org/}}, based on four centrality indices calculated on the Common Crawl web graph~\cite{CommonCrawl}. Depending on the index, these ranks vary widely even for very popular sites. Moreover, such centrality indices can be affected by manipulation~\cite{Gyongyi2006, Niu2015}.

In his analysis of DNS traffic from a Tor exit node, Sonntag~\cite{Sonntag2018} finds that popularity according to Alexa does not imply regular traffic over Tor, listing several domains with a good Alexa rank but that are barely seen in the DNS traffic. These conclusions confirm that different sources show a different view of popularity, and that the Alexa list may not be the most appropriate for all types of research (e.g.\ into Tor).

\section{Conclusion}
We find that 133 studies in recent main security conferences base their experiments on domains from commercial rankings of the `top' websites. However, the data sources and methods used to compile these rankings vary widely and their details are unknown, and we find that hidden properties and biases can skew research results. In particular, through an extensive evaluation of these rankings, we detect a recent unannounced change in the way Alexa composes its list: their data is only averaged over a single day, causing half of the list to change every day. Most probably, this unknowingly affected research results, and may continue to do so. However, other rankings exhibit similar problems: e.g.\ only 49\% of domains in Umbrella's list respond with HTTP status code 200, and Majestic's list, which Quad9 uses as a whitelist, has more than 2\,000 domains marked as malicious by Google Safe Browsing.

The reputational or commercial incentives in biasing the results of security studies, as well as the large trust placed in the validity of these rankings by researchers, as evidenced by only two studies putting their methods into question, makes these rankings an interesting target for adversarial manipulation. We develop techniques that exploit the pitfalls in every list by forging the data upon which domain rankings are based. Moreover, many of these methods bear an exceptionally low cost, both technically and in resources: we only needed to craft a single HTTP request to appear in Alexa's top million sites. This provides an avenue for manipulation at a very large scale, both in the rank that can be achieved and in the number of domains artificially inserted into the list. Adversaries can therefore sustain massive manipulation campaigns over time to have a significant impact on the rankings, and, as a consequence, on research and the society at large.

Ranking providers carry out few checks on their traffic data, as is apparent from our ability to insert nonexistent domains, further simplifying manipulation at scale. We outline several mitigation strategies, but cannot be assured that these will be implemented. Therefore, we introduce \listname, a new ranking based on combining the four existing lists, alongside the ability to filter out undesirable (e.g.\ unavailable or malicious) domains. These combined lists show much better stability over time, only changing by at most 0.6\% per day, and are much more resilient against manipulation, where even manipulating one list to reach the top 1\,000 only yields a rank of 100\,000 in our combined list. We offer an online service at \url{https://tranco-list.eu} to access these rankings in a reproducible manner, so that researchers can continue their evaluation with a more reliable and suitable set of domains. This helps them in assuring the validity, verifiability and reproducibility of their studies, making their conclusions about security on the Internet more accurate and well founded.

\section*{Acknowledgment}

The authors would like to thank Vera Rimmer, Davy Preuveneers and Quirin Scheitle for their valuable input. This research is partially funded by the Research Fund KU Leuven. Victor Le Pochat holds a PhD Fellowship of the Research Foundation - Flanders (FWO).

\IEEEtriggeratref{53}
\IEEEtriggercmd{\enlargethispage{-0.5875in}}

\bibliographystyle{IEEEtranS}
\bibliography{bibexport}

\begin{thebibliography}{10}
\providecommand{\url}[1]{#1}
\csname url@samestyle\endcsname
\providecommand{\newblock}{\relax}
\providecommand{\bibinfo}[2]{#2}
\providecommand{\BIBentrySTDinterwordspacing}{\spaceskip=0pt\relax}
\providecommand{\BIBentryALTinterwordstretchfactor}{4}
\providecommand{\BIBentryALTinterwordspacing}{\spaceskip=\fontdimen2\font plus
\BIBentryALTinterwordstretchfactor\fontdimen3\font minus
  \fontdimen4\font\relax}
\providecommand{\BIBforeignlanguage}[2]{{%
\expandafter\ifx\csname l@#1\endcsname\relax
\typeout{** WARNING: IEEEtranS.bst: No hyphenation pattern has been}%
\typeout{** loaded for the language `#1'. Using the pattern for}%
\typeout{** the default language instead.}%
\else
\language=\csname l@#1\endcsname
\fi
#2}}
\providecommand{\BIBdecl}{\relax}
\BIBdecl

\bibitem{alexaextension}
\BIBentryALTinterwordspacing
(2018, Nov.) {Alexa Traffic Rank} - {Chrome} {Web Store}. [Online]. Available:
  \url{https://chrome.google.com/webstore/detail/alexa-traffic-
  rank/cknebhggccemgcnbidipinkifmmegdel}
\BIBentrySTDinterwordspacing

\bibitem{AbuRajab2008a}
M.~Abu~Rajab, F.~Monrose, A.~Terzis, and N.~Provos, ``Peeking through the
  cloud: {DNS}-based estimation and its applications,'' in \emph{Proc.\ ACNS},
  2008, pp. 21--38.

\bibitem{Acar2014}
G.~Acar, C.~Eubank, S.~Englehardt, M.~Juarez, A.~Narayanan, and C.~Diaz, ``The
  web never forgets: Persistent tracking mechanisms in the wild,'' in
  \emph{Proc.\ CCS}, 2014, pp. 674--689.

\bibitem{Adamic2002a}
L.~A. Adamic and B.~A. Huberman, ``Zipf's law and the {Internet},''
  \emph{Glottometrics}, vol.~3, pp. 143--150, 2002.

\bibitem{Alexa5}
\BIBentryALTinterwordspacing
{Alexa Internet, Inc.} (2008, Dec.) Global top sites. [Online]. Available:
  \url{https://web.archive.org/web/20081216072512/http://
  www.alexa.com:80/site/ds/top_sites}
\BIBentrySTDinterwordspacing

\bibitem{Alexa3a}
\BIBentryALTinterwordspacing
------. (2016, Jan.) Does {Alexa} have a list of its top-ranked websites?
  Archived on April 4, 2016. [Online]. Available:
  \url{https://web.archive.org/web/20160404003433/https://
  support.alexa.com/hc/en-us/articles/200449834-Does-Alexa-
  have-a-list-of-its-top-ranked-websites-}
\BIBentrySTDinterwordspacing

\bibitem{Alexa3b}
\BIBentryALTinterwordspacing
------. (2017, Jan.) Does {Alexa} have a list of its top-ranked websites?
  Archived on March 11, 2017. [Online]. Available:
  \url{https://web.archive.org/web/20170311160137/https://
  support.alexa.com/hc/en-us/articles/200449834-Does-Alexa-
  have-a-list-of-its-top-ranked-websites-}
\BIBentrySTDinterwordspacing

\bibitem{Alexa1}
\BIBentryALTinterwordspacing
------. (2017, Nov.) How are {Alexa}'s traffic rankings determined? [Online].
  Available: \url{https://support.alexa.com/hc/en-us/articles/200449744}
\BIBentrySTDinterwordspacing

\bibitem{Alexa8}
\BIBentryALTinterwordspacing
------. (2018, May) What exactly is the {Alexa Traffic Panel}? [Online].
  Available: \url{https://support.alexa.com/hc/en-us/articles/200080859}
\BIBentrySTDinterwordspacing

\bibitem{Twitter2}
\BIBentryALTinterwordspacing
{Alexa Support (\texttt{@AlexaSupport})}. (2016, Nov.) Thanks to customer
  feedback, the top {1M} sites is temporarily available again. {W}e'll provide
  notice before updating the file in the future. [Online]. Available:
  \url{https://twitter.com/Alexa_Support/status/ 801167423726489600}
\BIBentrySTDinterwordspacing

\bibitem{Twitter1}
\BIBentryALTinterwordspacing
------. (2016, Nov.) Yes, the top 1m sites file has been retired. [Online].
  Available: \url{https://twitter.com/Alexa_Support/status/ 800755671784308736}
\BIBentrySTDinterwordspacing

\bibitem{Alexa6}
\BIBentryALTinterwordspacing
{Amazon Web Services, Inc.} (2018, Mar.) {Alexa Top Sites}. [Online].
  Available: \url{https://aws.amazon.com/alexa-top-sites/}
\BIBentrySTDinterwordspacing

\bibitem{Alexa9}
\BIBentryALTinterwordspacing
------. (2018, Aug.) {Alexa Web Information Service}. [Online]. Available:
  \url{https://aws.amazon.com/awis/}
\BIBentrySTDinterwordspacing

\bibitem{AWS1}
\BIBentryALTinterwordspacing
------. (2018, Apr.) {AWS IP} address ranges. [Online]. Available:
  \url{https://docs.aws.amazon.com/general/latest/gr/aws-ip- ranges.html/}
\BIBentrySTDinterwordspacing

\bibitem{Aydin2017}
\BIBentryALTinterwordspacing
Y.~Aydin. (2017, Oct.) \BIBforeignlanguage{French}{Étude nationale portant sur
  la sécurité de l’espace numérique français 2017}. Available in English
  at
  \url{https://www.economie.gouv.fr/files/2017_National_Study_Cybersecurity.pdf}.
  [Online]. Available: \url{https://www.economie.gouv.fr/files/
  2017_Etude_nationale_securite_numerique.pdf}
\BIBentrySTDinterwordspacing

\bibitem{Blundo2010}
C.~Blundo, S.~Cimato, and B.~Masucci, \emph{Secure Metering Schemes}.\hskip 1em
  plus 0.5em minus 0.4em\relax Springer US, 2010, pp. 1--32.

\bibitem{Borgolte2015}
K.~Borgolte, C.~Kruegel, and G.~Vigna, ``Meerkat: Detecting website defacements
  through image-based object recognition,'' in \emph{Proc.\ {USENIX} Security},
  2015, pp. 595--610.

\bibitem{ISPSellData}
\BIBentryALTinterwordspacing
J.~Brodkin. (2017, Mar.) How {ISPs} can sell your {Web} history --- and how to
  stop them. Ars Technica. [Online]. Available:
  \url{https://arstechnica.com/information-technology/2017/03/how-
  isps-can-sell-your-web-history-and-how-to-stop-them/}
\BIBentrySTDinterwordspacing

\bibitem{BuildWithCertify}
\BIBentryALTinterwordspacing
{BuiltWith Pty Ltd}. (2018, Sep.) Alexa {Certified} site metrics usage
  statistics. [Online]. Available:
  \url{https://trends.builtwith.com/analytics/Alexa-Certified-Site- Metrics}
\BIBentrySTDinterwordspacing

\bibitem{Cao2012}
Q.~Cao, M.~Sirivianos, X.~Yang, and T.~Pregueiro, ``Aiding the detection of
  fake accounts in large scale social online services,'' in \emph{Proc.\ NSDI},
  2012, pp. 197--210.

\bibitem{Umbrella1}
\BIBentryALTinterwordspacing
{Cisco Umbrella}. (2016) Umbrella popularity list. [Online]. Available:
  \url{https://s3-us-west-1.amazonaws.com/umbrella-static/ index.html}
\BIBentrySTDinterwordspacing

\bibitem{Clauset2009}
A.~Clauset, C.~R. Shalizi, and M.~E.~J. Newman, ``Power-law distributions in
  empirical data,'' \emph{SIAM Review}, vol.~51, no.~4, pp. 661--703, 2009.

\bibitem{CommonCrawl}
\BIBentryALTinterwordspacing
{Common Crawl Foundation}. {Common Crawl}. [Online]. Available:
  \url{https://commoncrawl.org/}
\BIBentrySTDinterwordspacing

\bibitem{Dittrich2012}
D.~Dittrich and E.~Kenneally, ``The {Menlo Report}: Ethical principles guiding
  information and communication technology research,'' U.S. Department of
  Homeland Security, Tech. Rep., Aug. 2012.

\bibitem{Doty2018}
\BIBentryALTinterwordspacing
N.~Doty, ``Mitigating browser fingerprinting in web specifications,'' W3C,
  {W3C} Editor's Draft, Jul. 2018. [Online]. Available:
  \url{https://w3c.github.io/fingerprinting-guidance/}
\BIBentrySTDinterwordspacing

\bibitem{Englehardt2016}
S.~Englehardt and A.~Narayanan, ``Online tracking: A 1-million-site measurement
  and analysis,'' in \emph{Proc.\ CCS}, 2016, pp. 1388--1401.

\bibitem{Felt2017}
A.~P. Felt, R.~Barnes, A.~King, C.~Palmer, C.~Bentzel, and P.~Tabriz,
  ``Measuring {HTTPS} adoption on the web,'' in \emph{Proc.\ {USENIX}
  Security}, 2017, pp. 1323--1338.

\bibitem{Fraenkel2014}
J.~Fraenkel and B.~Grofman, ``The {Borda} count and its real-world
  alternatives: Comparing scoring rules in {Nauru} and {Slovenia},''
  \emph{Australian Journal of Political Science}, vol.~49, no.~2, pp. 186--205,
  2014.

\bibitem{Quad9Majestic}
\BIBentryALTinterwordspacing
S.~Gallagher. (2017, Nov.) New “{Quad9}” {DNS} service blocks malicious
  domains for everyone. Ars Technica. [Online]. Available:
  \url{https://arstechnica.com/information-technology/2017/11/new-
  quad9-dns-service-blocks-malicious-domains-for-everyone/}
\BIBentrySTDinterwordspacing

\bibitem{STARTTLSArs}
\BIBentryALTinterwordspacing
D.~Goodin. (2015, Oct.) Don't count on {STARTTLS} to automatically encrypt your
  sensitive e-mails. Ars Technica. [Online]. Available:
  \url{https://arstechnica.com/information-technology/2015/10/dont-
  count-on-starttls-to-automatically-encrypt-your-sensitive-e- mails/}
\BIBentrySTDinterwordspacing

\bibitem{LogjamArs}
\BIBentryALTinterwordspacing
------. (2015, May) {HTTPS}-crippling attack threatens tens of thousands of
  {Web} and mail servers. Ars Technica. [Online]. Available:
  \url{https://arstechnica.com/information-technology/2015/05/
  https-crippling-attack-threatens-tens-of-thousands-of-web- and-mail-servers/}
\BIBentrySTDinterwordspacing

\bibitem{HTTPSArs}
\BIBentryALTinterwordspacing
------. (2016, Mar.) More than 11 million {HTTPS} websites imperiled by new
  decryption attack. Ars Technica. [Online]. Available:
  \url{https://arstechnica.com/information-technology/2016/03/more-
  than-13-million-https-websites-imperiled-by-new-decryption- attack/}
\BIBentrySTDinterwordspacing

\bibitem{SafeBrowsing}
\BIBentryALTinterwordspacing
{Google Inc.} Safe browsing. [Online]. Available:
  \url{https://safebrowsing.google.com/}
\BIBentrySTDinterwordspacing

\bibitem{CrUX1}
\BIBentryALTinterwordspacing
{Google, Inc.} (2018, Jan.) {Chrome User Experience Report}. [Online].
  Available: \url{https://developers.google.com/web/tools/chrome-user-
  experience-report/}
\BIBentrySTDinterwordspacing

\bibitem{Gyongyi2006}
Z.~Gyöngyi, P.~Berkhin, H.~Garcia-Molina, and J.~Pedersen, ``Link spam
  detection based on mass estimation,'' in \emph{Proc.\ VLDB}, 2006, pp.
  439--450.

\bibitem{NETRESEC}
\BIBentryALTinterwordspacing
E.~Hjelmvik. (2017, Apr.) Domain whitelist benchmark: {Alexa} vs {Umbrella}.
  NETRESEC. [Online]. Available:
  \url{https://www.netresec.com/?page=Blog&month=2017-04&
  post=Domain-Whitelist-Benchmark%3A-Alexa-vs-Umbrella}
\BIBentrySTDinterwordspacing

\bibitem{Umbrella2}
\BIBentryALTinterwordspacing
D.~Hubbard. (2016, Dec.) {Cisco Umbrella} 1 million. [Online]. Available:
  \url{https://umbrella.cisco.com/blog/2016/12/14/cisco-umbrella-1- million/}
\BIBentrySTDinterwordspacing

\bibitem{Invernizzi2016}
L.~Invernizzi, K.~Thomas, A.~Kapravelos, O.~Comanescu, J.~Picod, and
  E.~Bursztein, ``Cloak of visibility: Detecting when machines browse a
  different web,'' in \emph{Proc.\ SP}, 2016, pp. 743--758.

\bibitem{Majestic2}
\BIBentryALTinterwordspacing
D.~Jones. (2012, Oct.) {Majestic Million CSV} now free for all, daily.
  [Online]. Available:
  \url{https://blog.majestic.com/development/majestic-million-csv- daily/}
\BIBentrySTDinterwordspacing

\bibitem{Juba2015}
B.~Juba, C.~Musco, F.~Long, S.~Sidiroglou-Douskos, and M.~Rinard, ``Principled
  sampling for anomaly detection,'' in \emph{Proc.\ NDSS}, 2015.

\bibitem{Karami2016}
M.~Karami, Y.~Park, and D.~McCoy, ``Stress testing the booters: Understanding
  and undermining the business of {DDoS} services,'' in \emph{Proc.\ WWW},
  2016, pp. 1033--1043.

\bibitem{Kumar2017}
D.~Kumar, Z.~Ma, Z.~Durumeric, A.~Mirian, J.~Mason, J.~A. Halderman, and
  M.~Bailey, ``Security challenges in an increasingly tangled web,'' in
  \emph{Proc.\ WWW}, 2017, pp. 677--684.

\bibitem{Lever2016}
C.~Lever, R.~Walls, Y.~Nadji, D.~Dagon, P.~McDaniel, and M.~Antonakakis,
  ``{Domain-Z}: 28 registrations later. {M}easuring the exploitation of
  residual trust in domains,'' in \emph{Proc.\ SP}, 2016, pp. 691--706.

\bibitem{Libert2015}
T.~Libert, ``Exposing the hidden web: An analysis of third-party {HTTP}
  requests on 1 million websites,'' \emph{International Journal of
  Communication}, vol.~9, pp. 3544--3561, Oct. 2015.

\bibitem{Libert2018}
------, ``An automated approach to auditing disclosure of third-party data
  collection in website privacy policies,'' in \emph{Proc.\ WWW}, 2018, pp.
  207--216.

\bibitem{Lo2006}
B.~W.~N. Lo and R.~S. Sedhain, ``How reliable are website rankings?
  {Implications} for e-business advertising and {Internet} search,''
  \emph{Issues in Information Systems}, vol. VII, no.~2, pp. 233--238, 2006.

\bibitem{Umbrella3}
\BIBentryALTinterwordspacing
O.~Lystrup. (2016, Dec.) Cisco {Umbrella} releases free top 1 million sites
  list. [Online]. Available:
  \url{https://medium.com/cisco-shifted/cisco-umbrella-releases-
  free-top-1-million-sites-list-8497fba58efe}
\BIBentrySTDinterwordspacing

\bibitem{Majestic1}
\BIBentryALTinterwordspacing
{Majestic-12 Ltd.} Frequently asked questions. [Online]. Available:
  \url{https://majestic.com/support/faq}
\BIBentrySTDinterwordspacing

\bibitem{Majestic5}
\BIBentryALTinterwordspacing
------. (2018, Apr.) Majestic launch a bigger fresh index. [Online]. Available:
  \url{https://blog.majestic.com/company/majestic-launch-a-bigger-
  fresh-index/}
\BIBentrySTDinterwordspacing

\bibitem{Mertens2017}
\BIBentryALTinterwordspacing
X.~Mertens. (2017, Apr.) Whitelists: The holy grail of attackers. {SANS}
  Internet Storm Center. [Online]. Available:
  \url{https://isc.sans.edu/forums/diary/
  Whitelists+The+Holy+Grail+of+Attackers/22262/}
\BIBentrySTDinterwordspacing

\bibitem{Metwally2007}
A.~Metwally, D.~Agrawal, A.~E. Abbad, and Q.~Zheng, ``On hit inflation
  techniques and detection in streams of web advertising networks,'' in
  \emph{Proc.\ ICDCS}, 2007, pp. 52--52.

\bibitem{Meusel2015}
R.~Meusel, S.~Vigna, O.~Lehmberg, and C.~Bizer, ``The graph structure in the
  {Web} -- analyzed on different aggregation levels,'' \emph{The Journal of Web
  Science}, vol.~1, no.~1, pp. 33--47, 2015.

\bibitem{Moura2017}
G.~C.~M. Moura, M.~Müller, M.~Davids, M.~Wullink, and C.~Hesselman, ``Domain
  names abuse and {TLDs}: From monetization towards mitigation,'' in
  \emph{Proc.\ IM}, May 2017, pp. 1077--1082.

\bibitem{Napoli2014}
P.~M. Napoli, P.~J. Lavrakas, and M.~Callegaro, ``Internet and mobile ratings
  panels,'' in \emph{Online Panel Research: A Data Quality Perspective}.\hskip
  1em plus 0.5em minus 0.4em\relax Wiley-Blackwell, 2014, ch.~17, pp. 387--407.

\bibitem{Nikiforakis2012}
N.~Nikiforakis, L.~Invernizzi, A.~Kapravelos, S.~{Van Acker}, W.~Joosen,
  C.~Kruegel, F.~Piessens, and G.~Vigna, ``You are what you include:
  Large-scale evaluation of remote {JavaScript} inclusions,'' in \emph{Proc.\
  CCS}, 2012, pp. 736--747.

\bibitem{Niu2015}
Q.~Niu, A.~Zeng, Y.~Fan, and Z.~Di, ``Robustness of centrality measures against
  network manipulation,'' \emph{Physica A: Statistical Mechanics and its
  Applications}, vol. 438, pp. 124--131, 2015.

\bibitem{OpenDNSSystem}
\BIBentryALTinterwordspacing
{OpenDNS}. {OpenDNS System}. [Online]. Available:
  \url{https://system.opendns.com/}
\BIBentrySTDinterwordspacing

\bibitem{Partridge2016}
C.~Partridge and M.~Allman, ``Ethical considerations in network measurement
  papers,'' \emph{Communications of the ACM}, vol.~59, no.~10, pp. 58--64, Sep.
  2016.

\bibitem{Pavlo2011}
\BIBentryALTinterwordspacing
A.~Pavlo and N.~Shi, ``Graffiti networks: {A} subversive, {Internet}-scale file
  sharing model,'' \emph{ArXiv e-prints}, 2011. [Online]. Available:
  \url{http://arxiv.org/abs/1101.0350}
\BIBentrySTDinterwordspacing

\bibitem{Quantcast3}
\BIBentryALTinterwordspacing
{Quantcast}. (2007, Jul.) Open internet ratings service. [Online]. Available:
  \url{https://web.archive.org/web/20070705200342/http:// www.quantcast.com/}
\BIBentrySTDinterwordspacing

\bibitem{Rimmer2018}
V.~Rimmer, D.~Preuveneers, M.~Juarez, T.~Van~Goethem, and W.~Joosen,
  ``Automated website fingerprinting through deep learning,'' in \emph{Proc.\
  NDSS}, 2018.

\bibitem{Scheitle2018a}
Q.~Scheitle, O.~Hohlfeld, J.~Gamba, J.~Jelten, T.~Zimmermann, S.~D. Strowes,
  and N.~Vallina-Rodriguez, ``A long way to the top: Significance, structure,
  and stability of {Internet} top lists,'' in \emph{Proc.\ IMC}, 2018, pp.
  478--493.

\bibitem{ExpiredDomains}
\BIBentryALTinterwordspacing
M.~Schmidt. (2018, Aug.) Expired domains. [Online]. Available:
  \url{https://www.expireddomains.net}
\BIBentrySTDinterwordspacing

\bibitem{Quantcast1}
\BIBentryALTinterwordspacing
S.~Simpson. (2018, Jan.) For sites that are not {Quantified}, it says ``data is
  estimated.'' {What} does this mean? [Online]. Available:
  \url{https://quantcast.zendesk.com/hc/en-us/articles/ 115013961667}
\BIBentrySTDinterwordspacing

\bibitem{Quantcast2}
\BIBentryALTinterwordspacing
------. (2018, Jan.) What is the date range for the traffic numbers on your
  site? [Online]. Available:
  \url{https://quantcast.zendesk.com/hc/en-us/articles/ 115013961687}
\BIBentrySTDinterwordspacing

\bibitem{Sonntag2018}
M.~Sonntag, ``{DNS} traffic of a {Tor} exit node - an analysis,'' in
  \emph{Proc.\ SpaCCS}, 2018, pp. 33--45.

\bibitem{Tajalizadehkhoob2016}
S.~Tajalizadehkhoob, M.~Korczy{\'{n}}ski, A.~Noroozian, C.~Gañán, and M.~van
  Eeten, ``Apples, oranges and hosting providers: Heterogeneity and security in
  the hosting market,'' in \emph{Proc.\ NOMS}, 2016, pp. 289--297.

\bibitem{MozillaSymantec}
\BIBentryALTinterwordspacing
W.~Thayer. (2018, Oct.) Delaying further {Symantec} {TLS} certificate distrust.
  Mozilla Foundation. [Online]. Available:
  \url{https://blog.mozilla.org/security/2018/10/10/delaying-
  further-symantec-tls-certificate-distrust/}
\BIBentrySTDinterwordspacing

\bibitem{TorRelays}
\BIBentryALTinterwordspacing
{The Tor Project}. (2018, Apr.) Number or relays with relay flags assigned.
  [Online]. Available:
  \url{https://metrics.torproject.org/relayflags.html?start=2017-
  01-01&end=2018-04-26&flag=Exit}
\BIBentrySTDinterwordspacing

\bibitem{LogjamWSJ}
\BIBentryALTinterwordspacing
J.~Valentino-DeVries. (2015, May) New computer bug exposes broad security
  flaws. The Wall Street Journal. [Online]. Available:
  \url{https://www.wsj.com/articles/new-computer-bug-exposes-broad-
  security-flaws-1432076565}
\BIBentrySTDinterwordspacing

\bibitem{Vissers2015a}
T.~Vissers, W.~Joosen, and N.~Nikiforakis, ``Parking sensors: Analyzing and
  detecting parked domains,'' in \emph{Proc.\ NDSS}, 2015.

\bibitem{Webber2010a}
W.~Webber, A.~Moffat, and J.~Zobel, ``A similarity measure for indefinite
  rankings,'' \emph{ACM Transactions on Information Systems}, vol.~28, no.~4,
  pp. 1--38, Nov. 2010.

\bibitem{Alexa2}
\BIBentryALTinterwordspacing
J.~Yesbeck. (2014, Oct.) Your top questions about {Alexa} data and ranks,
  answered. [Online]. Available:
  \url{https://blog.alexa.com/top-questions-about-alexa-answered/}
\BIBentrySTDinterwordspacing

\bibitem{Zhang2018}
X.~Zhang, X.~Wang, X.~Bai, Y.~Zhang, and X.~Wang, ``{OS}-level side channels
  without procfs: Exploring cross-app information leakage on {iOS},'' in
  \emph{Proc.\ NDSS}, 2018.

\bibitem{Zimmeck2017}
S.~Zimmeck, J.~S. Li, H.~Kim, S.~M. Bellovin, and T.~Jebara, ``A privacy
  analysis of cross-device tracking,'' in \emph{Proc.\ {USENIX} Security},
  2017, pp. 1391--1408.

\end{thebibliography}

\end{document}